\documentclass[useAMS,usenatbib]{mn2e}
\usepackage{graphicx}

\title[CORS Baade--Wesselink method in the Walraven system.
  ]{CORS Baade--Wesselink method in the Walraven photometric system: the 
  Period--Radius and the Period--Luminosity relation of Classical Cepheids.}
\author[R. Molinaro et al.]
  {R. Molinaro$^{1}$\thanks{E--mail: molinaro@na.astro.it},
   V. Ripepi$^{1}$, M. Marconi$^{1}$, G. Bono$^{2,3}$,
   J. Lub$^{4}$, S. Pedicelli$^{2}$, J.W. Pel$^{5}$\\
$^{1}$INAF--Osservatorio Astronomico di Capodimonte, Via Moiariello 16,
  80131, Napoli, Italy\\
$^{2}$Dipartimento di Fisica-- Univ. di Roma Tor Vergata, via della Ricerca Scientifica 1, 00133 Roma, Italy\\
$^{3}$ INAF--Osservatorio Astronomico di Roma, Via Frascati 33, 00040, Monte Porzio Catone, Italy\\
$^{4}$ Leiden Observatory, Leiden University, PO Box 9513, 2300 RA Leiden,
The Netherlands\\
$^{5}$ Kapteyn Institute, University of Groningen, PO Box 800, 9700 AV
Groningen, The Netehrlands\\ 
}
\begin{document}

\date{Accepted: Received:}

\pagerange{\pageref{firstpage}--\pageref{lastpage}} \pubyear{2010}

\maketitle

\label{firstpage}

\begin{abstract}
We present a new derivation of the CORS Baade--Wesselink method in the
Walraven photometric system.  We solved the complete Baade--Wesselink
equation by calibrating the surface brightness function with a recent grid
of atmosphere models. The new approach was adopted to estimate the mean
radii of a sample of Galactic Cepheids for which are available precise
light curves  in the Walraven bands. Current radii agree, within the
errors, quite well  with Cepheid radii based on  recent optical and
near--infrared interferometric measurements.

We also tested the impact of the projection factor on the Period--Radius
relation using  two different values ($p=1.36$, $p=1.27$) that
bracket the estimates available in the literature. We found that the
agreement of our Period--Radius relation with similar empirical and
theoretical Period--Radius relations in the recent literature, improves
by changing the projection factor from $p=1.36$ to $p=1.27$. Our
Period--Radius relation is $\log R=(0.75\pm 0.03)\log P+(1.10\pm 0.03)$,
with a $rms=0.03$ dex. 

Thanks to accurate estimates of the effective temperature of the selected
Cepheids, we also derived the Period--Luminosity relation in the
V band and we found $M_V=(-2.78\pm 0.11)\log P+(-1.42\pm 0.11)$ with
$rms=0.13$ mag, for $p=1.27$. It agrees quite well with recent results in
the literature, while the relation for $p=1.36$ deviates
by more than $2\sigma$. We conclude that, even taking into account the
intrinsic dispersion of the obtained Period--Luminosity relations, that is roughly of
the same order of magnitude as the effect of the projection factor,
the results of this paper seem to favour the value $p=1.27$.  
\end{abstract}


\section{Introduction}
Classical Cepheids are fundamental standard candles to determine both 
Galactic and extragalactic distances. They obey  well defined 
optical and Near--Infrared  Period--Luminosity  and 
Period--Luminosity--Colour  relations. Thanks to the unprecedented 
spatial resolution of the Hubble Space Telescope, Cepheids can be 
used as primary distance indicators up to the Virgo cluster 
($\approx$ 25 Mpc). This means that they can also be adopted to provide 
robust absolute calibrations of secondary distance indicators, and in turn 
to estimate the value of the Hubble constant \citep[see e.g.][]{fm01}. 

The accuracy reached in the determination of cosmological parameters 
using the Cepheid distance scale relies on the knowledge of the 
physical mechanisms that govern their radial oscillations and on 
their dependence on intrinsic parameters (luminosity, mass, radius, 
chemical composition; see e.g. \citet{bms99, bms00, bcm02, kw06}, and
references therein).
In particular, the estimate of Cepheid radii and of the effective 
temperature (using mean intrinsic colours) allows us to constrain their 
intrinsic luminosity. Furthermore, the 
Cepheid mass can be estimated by adopting a Period--Mass--Radius relation   
\citep[see e.g.][and references therein]{bg01}, and in turn to provide 
independent constraints on the mass-luminosity relations.     

The methods currently used to derive Cepheid radii from photometric
and spectroscopic data (radial velocity) are based on the classical
Baade--Wesselink  technique \citep{w46}. Among them, the surface 
brightness technique \citep{be76} and the CORS method \citep{co81} 
rely on solid physical bases. The reader interested in a detailed 
discussion concerning  the different Baade--Wesselink methods is 
referred to \citet{g87}. 
The use of interferometric methods and the measurement of parallaxes
provide a model-independent method for radius measurements. However,  
this approach has only been applied to a very limited sample of bright 
nearby Cepheids \citep{na00, lc02, kb04, dj09}. 
On the the other hand, the Baade--Wesselink methods can be applied to large samples 
of Cepheids and it has been recently extended to Magellanic Cepheids 
\citep{sg05}. However, the precision of Baade--Wesselink distances strongly 
depends on the quality of photometric and radial velocity data.

In a previous paper \citep{rr04}, we adapted the CORS method to
derive Cepheid radii observed in the Str\"omgren system. However the 
model atmospheres adopted in the quoted investigation to estimate the 
surface brightness did not allow us to include Cepheids with periods 
longer than 12--13 days. The cutoff in the long period range limited the
accuracy of the Period--Radius relation.

Recently, \citet{pb09} provided new calibrations based on a new
release of model atmospheres by \citet{c99} for the Walraven
photometry of Galactic Cepheids originally collected by \citet{p78}.
The Walraven system includes five bands, (V, B, L, U, W), designed to
measure the features of hydrogen spectrum. Three of them have the
central wavelength in the ultraviolet region \citep[see][for more
details about this photometric system]{lp77,pl07}. In particular the
central wavelength of the Walraven filters are the following:
 $\lambda_V=5405\AA,\,\lambda_B=4280\AA,\,\lambda_L=3825\AA,\,\lambda_U=3630\AA,\,\lambda_W=3240\AA$.  

In this investigation we show that this new calibration of the VBLUW Cepheid 
time series data by \citet{pb09} allows us to derive precise stellar parameters 
(surface gravity and effective temperature) at fixed chemical composition
and to calibrate with high accuracy the surface brightness function,
a fundamental ingredient for the application of the Baade--Wesselink method.
This opportunity stimulated us to apply the CORS version of the
Baade--Wesselink technique as formulated  
by \citet{ob85} and \citet{rr04}, to derive reliable radii for a sample of 
Cepheids having periods from a few days up to $\sim 40$ days and to obtain 
more accurate Period--Radius and Period--Luminosity relations.
 
The layout of the paper is the following: the original CORS
method is introduced in Sec.~\ref{sec-cors}. The grids of atmosphere 
models and the calibration of the surface brightness in the Walraven 
system are described in Sec.~\ref{sec-cors-modif}. In Sec.~\ref{sec-data} 
we describe the adopted photometric and spectroscopic data, while the 
estimate of the Cepheids radii and of the Period--Radius relation are 
discussed in Secs.~\ref{sec-radii} and  \ref{sec-pr}, respectively. 
In Sec.~\ref{sec-pl} we discuss the derivation of the
Period--Luminosity relation.
Finally, Sec.~\ref{sec-conclusion} summarizes the main results of this 
investigation.

\section{The CORS method}
\label{sec-cors}
The original CORS method \citep{co81} is a variant of the classical Baade--Wesselink 
technique \citep{w46} and was developed to derive the radii of pulsating stars. It relies 
on the surface brightness function:
\begin{equation}
S_V= m_V + 5\log \alpha
\label{eq-sb}
\end{equation}
where $m_V$ is the apparent visual magnitude and $\alpha$ is the angular
diameter of the star. For a variable star Eq. (\ref{eq-sb}) can be written 
for each phase $(\phi)$ along the pulsational cycle. By differentiating the 
above equation with respect to the phase, by multiplying the result with 
a colour index $(C_{ij}$, where $i$,$j$ are two generic photometric bands) 
and by integrating over the entire cycle, one can write:

\begin{equation}
q\int_{0}^{1}\!\ln
\Big\{R_0-pP\int_{\phi_0}^{\phi}\!v(\phi')d\phi'\Big\}C'_{ij}d\phi-B+\Delta
B=0
\label{eq-cors}
\end{equation}

where $q=\frac{5}{\ln 10}$, $P$ is the period, $v$ is the radial velocity 
and $p$ is the radial velocity projection factor. The other two terms, $B$ 
and $\Delta B$, are the following:
\begin{eqnarray}
B=\int_{0}^{1}C_{ij}(\phi)m'_V(\phi)d\phi \hspace{10pt} \\
\Delta B=\int_{0}^{1}C_{ij}(\phi)S'_V(\phi)d\phi \; . \label{eq-db}
\end{eqnarray}

The projection factor $p$ correlates radial and pulsation velocity, i.e.  
$R'(\phi)=-p\cdot P \cdot v(\phi)$. The uncertainty affecting the estimates 
of this parameter is still the main source of systematic errors in the Baade--Wesselink 
method. We shall discuss the $p$ value in Sec.~\ref{sec-pfactor}.

By solving Eq. (\ref{eq-cors}) we estimate the value of the radius $R_0$ at
an arbitrary phase $\phi_0$, while the radius at any phase $\phi$ along 
the pulsational cycle can be evaluated by integrating the radial velocity 
curve between $\phi_0$ and $\phi$. The term $B$ present in Eq. (\ref{eq-cors}) 
can be easily determined using colour and magnitude ($B$), while the term $\Delta B$
is not directly connected to observational data, but approximates the
area of the loop performed by variable stars in the colour-colour plane. 

Typically the $\Delta B$ term has a small value \citep[$10^{-3}-10^{-4}$,][]{ob85} 
and in the original Baade--Wesselink method it is neglected \citep[see][]{co81}. However, the 
Cepheid radii estimated by including the $\Delta B$ term in the Eq. (\ref{eq-cors}) 
are more accurate than those based on the original Baade--Wesselink method \citep{sr81,rb97,rr00,rr04}.  

In the next section we describe how to calibrate the surface brightness function 
to estimate the term $\Delta B$ from Eq. (\ref{eq-db}).

\section{The modified CORS method}
\label{sec-cors-modif}
\subsection{The grids of atmosphere models}
In order to calibrate the surface brightness function, we assume
the validity of the quasi--static approximation (QSA)\footnote[1]{The
  quasi--static approximation assumes that the atmosphere of the
  pulsating star can be described, at any time, by a classical
  hydrostatic, plane parallel model in radiative/convective
  equilibrium and in local thermodynamic equilibrium, identified by the
  effective temperature $(T_{eff})$ and by the effective gravity
  $(g_{eff}=\frac{GM}{R^2}+\frac{d^2R}{dt^2})$.} for the Cepheid
atmosphere \citep[see also][]{ob85}. Under this hypothesis, any 
photometric quantity can be expressed as a function of effective temperature and 
gravity $(T_{eff}$, $g_{eff})$. Then, we can write:
\begin{equation}
S_V=S_V(T_{eff},\,g_{eff}) 
\end{equation}
for the surface brightness, and:
\begin{eqnarray}
\label{eq-direct}
C_{ij}=C_{ij}(T_{eff},\,g_{eff}) \\
C_{kl}=C_{kl}(T_{eff},\,g_{eff}) \nonumber
\end{eqnarray}
for two intrinsic colour indices.
If the determinant of the Jacobian $J(C_{ij},
\,C_{kl}|T_{eff},\,g_{eff})\neq 0$ the last two equations can be inverted and we obtain: 
\begin{eqnarray}
\label{eq-inverse}
T_{eff}=T_{eff}(C_{ij},\,C_{kl}) \\
g_{eff}=g_{eff}(C_{ij},\,C_{kl}) \nonumber
\end{eqnarray}
and we can also write the surface brightness as a function of the two colour indices:
\begin{equation}
S_V=S_V(C_{ij},\,C_{kl}) \; .
\label{eq-sb-colour}
\end{equation}

\begin{figure}
\includegraphics[width=84mm]{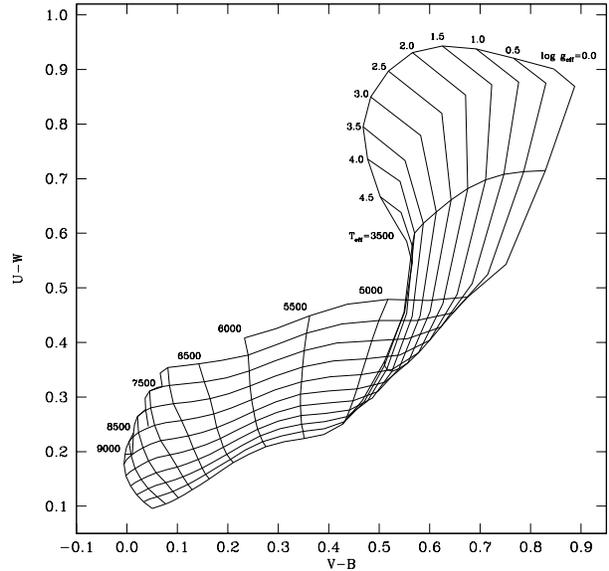}
\caption{The grids of model atmosphere plotted in the plane of $(V-B)$,
  $(U-W)$ colours: lines of constant temperature and constant gravity are
  plotted.} 
\label{fig-griglia}
\end{figure}

In general, the invertibility condition is not valid over the entire 
parameter space, since the same pair of colours trace different pairs of 
gravity and temperature ($g_{eff},\,T_{eff}$). Figure
~\ref{fig-griglia} shows the colour degeneracy present in the grid of
atmosphere model provided by \citet{c99} ({\it
  http://wwwuser.oat.ts.astro.it/castelli/colors/vbluw.html}), for the
Walraven two-colour diagram $(V-B)$ vs $(U-W)$\footnote[2]{We
  remember 
  that in the Walraven photometric system the magnitude in a generic
  band X is defined without the $-2.5$ factor: $m_X=\log F_X$. As a
  consequence, the colour is defined as the difference between the
  magnitude at longer and at shorter wavelength.}.  

However, a local solution can be found, provided that an appropriate choice 
of the colours $C_{ij}$ and $C_{kl}$ and of the range in colour is made.
By using the quoted colours $(V-B)$ and $(U-W)$, we succeeded in inverting 
Eq. (\ref{eq-direct}) for the range of parameters typical of Cepheids, 
i.e.  $0.0\le \log g_{eff} \le 3.0$ dex and $4500\le T_{eff} \le 7250$ K. 

\begin{figure}
\includegraphics[width=84mm]{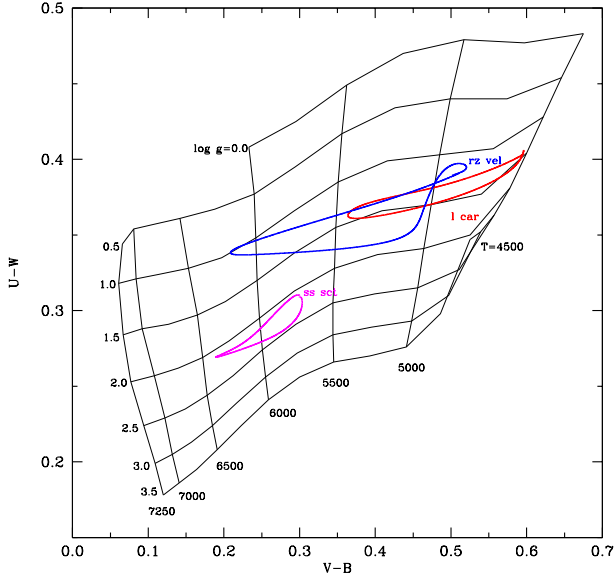}
\caption{Colour--colour loops for three selected Cepheids plotted onto the
  theoretical grids. The three Cepheids are SS Sct ($P=3.7$ days),
  RZ Vel ($P=20.4$ days) and l Car ($P=35.5$ days).} 
\label{fig-griglia-loop}
\end{figure}

To verify that real Cepheids are located inside the theoretical grid, 
we plotted the loops, obtained by fitting the observed colours of selected 
Cepheids, in the quoted colour--colour plane. The result for three Cepheids 
with short-, medium- and long-period, is shown in Fig.~\ref{fig-griglia-loop}. 
A glance at the data plotted in this figure shows that the three loops performed 
by actual Cepheids are located inside the limits of the predicted grid.  

\citet{lp77} and \citet{p78} pointed out that the flux in the $W$ band was measured 
with sufficient accuracy only for the brightest Cepheids due to the
intrinsic faintness of Cepheid fluxes in the near-ultraviolet, but
also because the telescope mirror reflectivity 
in this band was deteriorated at the epoch of observations. Therefore, we decided 
to perform a further test by using the pair of colours $(V-B)$ and $(L-U)$, even 
if the curves of constant temperature and gravity are not orthogonal as in the plane 
with the $(U-W)$ colour. 
Accurate solutions can be obtained in the range $5000\le T_{eff}<7500$ K and 
$0.0\le \log g_{eff} \le 4.00$ dex. However, the loops of Cepheids with 
periods longer than 11 days fall outside the grid of models. This means that the 
structural parameters of Cepheids and the surface brightness function are less 
accurate using this colour--colour plane. Nevertheless, the ratios of
the Cepheid radii with $P<11$ days, estimated using the two different
colour--colour planes, show that the differences are smaller than
$10\%$. On the basis of this evidence, we decided to focus our
attention on the $(V-B)-(U-W)$ plane (see Appendix ~\ref{appB} for
more details).   
   
\subsection{Derivation of $S_V$}
The availability of a region across the theoretical grid covering the
locus occupied by the Cepheid colour--colour loops and the validity of
the invertibility condition shown in the above section provide us
with an instrument to write effective temperature and gravity as
a function of the two colour indices and, finally to calibrate the 
surface brightness through Eq. (\ref{eq-sb-colour}). 

\begin{figure}
\includegraphics[angle=270,width=90mm]{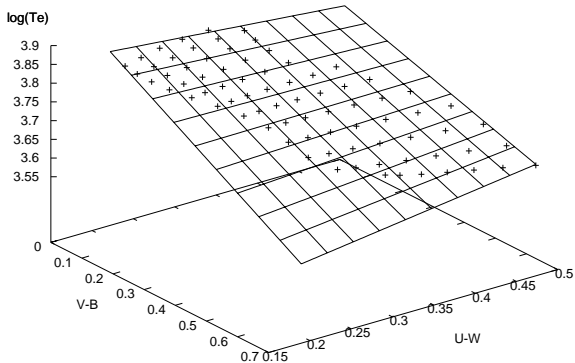}
\caption{The surface representing $\log T_{eff}$ as a function of
  $(V-B)$ and $(U-W)$, obtained from the polynomial fit of the atmosphere 
  models (crosses) from \citet{c99}.} 
\label{fig-fit-t}
\end{figure}

\begin{figure}
\includegraphics[angle=270,width=90mm]{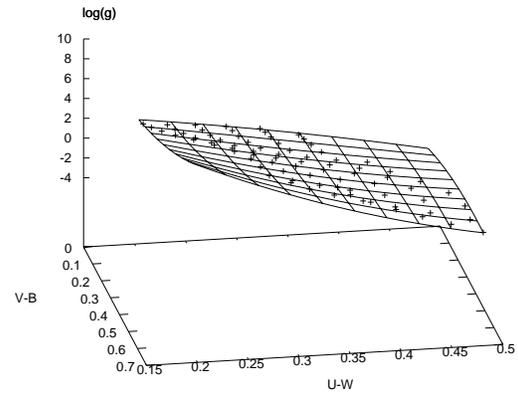}
\caption{The same as Fig.~\ref{fig-fit-t}, but for $\log g_{eff}$.}  
\label{fig-fit-g}
\end{figure}

As a first step, we have fitted the theoretical grid shown
in Fig.~\ref{fig-griglia-loop} by means of polynomials, in order to
express temperature and gravity as a function of $(V-B)$ and $(U-W)$
colours, according to Eq. (\ref{eq-inverse}) (see Appendix ~\ref{appA}
for details about the fitted equations). The surfaces $\log T_{eff}=\log
T_{eff}(V-B,\,U-W)$ and $\log g_{eff}=\log g_{eff}(V-B,\,U-W)$ are
shown in Fig.~\ref{fig-fit-t}, ~\ref{fig-fit-g}. The rms around the
fitted equations amount to 0.0033 dex and 0.15 dex for $\log T_{eff}$
and $\log g_{eff}$, respectively. 

\begin{figure}
\includegraphics[angle=270,width=90mm]{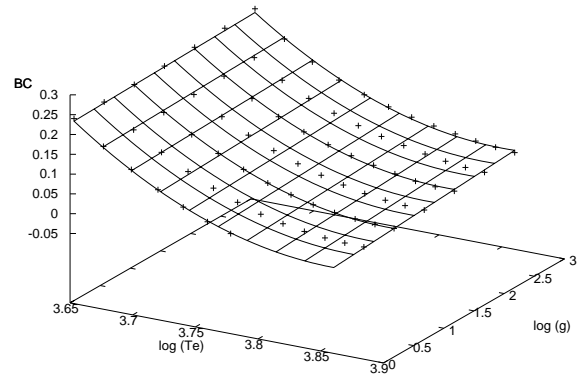}
\caption{The surface representing the Bolometric Correction, $BC$, as
  a function of $\log T_{eff}$ and $\log g_{eff}$, obtained from the
  polynomial fit of atmosphere models (crosses) from \citet{c99}.} 
\label{fig-fit-BC}
\end{figure}

Following the Walraven convention for magnitudes, the surface
brightness can be derived in the following way:
\begin{equation}
S_V=const. +4\log T_{eff} -BC(T_{eff},\,g_{eff})
\end{equation}

where $BC$ is the Bolometric Correction calculated as a function of
$T_{eff}$ and $g_{eff}$ through a polynomial fit (see Appendix
~\ref{appA}). A plot of the fit of $BC$ as a function of $\log T_{eff}$
and $\log g_{eff}$ is shown in Fig.~\ref{fig-fit-BC}.

The procedure described here allows us to calculate the term $\Delta
B$ in the complete CORS formulation (Eq.~\ref{eq-cors}).

\subsection{Test with spectroscopic and interferometric data.}

In the previous section we have described the procedure followed to derive
the physical parameters of Cepheids starting from the measured
colours. In order to test the accuracy of the fitted relations
(Eq.~\ref{eq-inverse}), we have performed a comparison of the effective
temperature and gravity derived from the fitted relations for selected  
Cepheids in our sample with the results obtained by other authors using  
an independent spectroscopic approach. 

\begin{figure}
\includegraphics[width=84mm]{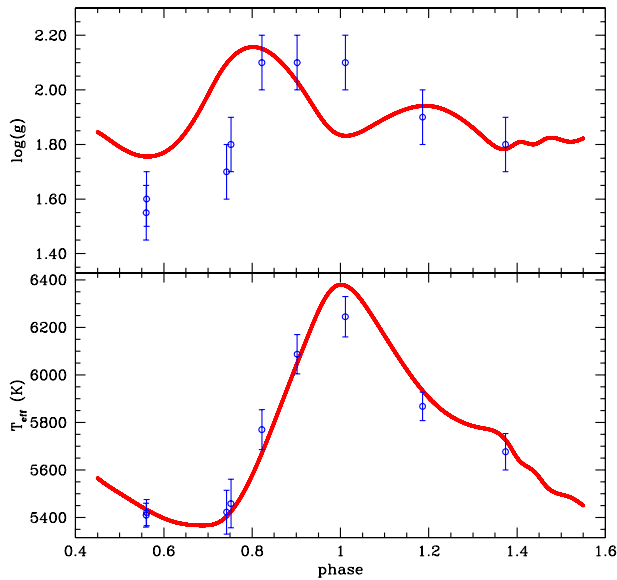}
\caption{Effective temperature (bottom panel) and gravity (upper
  panel), obtained by fitting the grid of model atmosphere, are
  plotted as function of the pulsational phase for the star U Sgr
  ($P=7.9$ days). The open blue circles with the associated error bars
  represent the values of the physical parameters obtained
  spectroscopically by \citet{la04}.}   
\label{fig-test}
\end{figure}

As an example, in Fig. ~\ref{fig-test} we have plotted the values of
effective temperature and gravity obtained with our equations for the
pulsational cycle of the Cepheid U Sgr. In the same figure we also 
overplotted the values of the physical parameters obtained
spectroscopically by \citet{la04} using the line depth ratios. 
We note that the curve of effective temperature follows the spectroscopic 
data with great accuracy. The agreement is less satisfactory for the curve 
of gravity, probably as a consequence of the cited problems with the $W$ band. 
Similar results are obtained for another dozen of stars for which the comparison 
with spectroscopic data was possible. We can conclude, that our approximation 
for the effective temperature is quite accurate, with mean relative
differences of $0.4\%$. On the 
other hand, we can reproduce the effective gravity, $\log g_{eff}$, with mean relative differences 
of $4\%$. However, the larger error affecting the estimate of the surface gravity 
is not a thorny problem in the estimate of Cepheid radii, because the gravity 
enters only in the determination of the $\Delta$B term, through the contribution 
of the BC to the surface brightness ($S_V$). The $\Delta$B term contributes at most 
with a 5\% to the radius, therefore, we can conclude that even a significant uncertainty 
in our determination of the gravity has a minimal impact on the determination of Cepheid 
radii.  
     
\begin{figure}
\includegraphics[width=84mm]{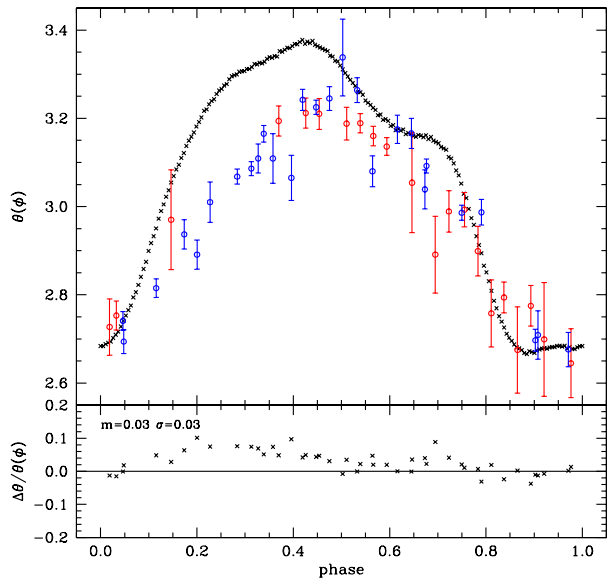}
\caption{Top panel: the curve of angular diameter $\theta$ (in mas) of l Car,
  obtained from the surface brightness calibration using
  eq.~\ref{eq-ang-diam} (black points), is plotted with the
  interferometric measurements obtained by \citet{kb04} (red points) and
  by \citet{dj09} (blue points). 
  Bottom  panel: difference between the two curves the mean and standard
  deviation of such a difference is labelled.}   
\label{fig-test2}
\end{figure}

As a further test of the accuracy of our approximation for the surface
brightness $S_V$, we have compared the angular diameter $\theta$ of l
Car (which is the only single Cepheid for which this comparison is
possible) obtained  through interferometric measurements \citep{kb04, dj09} with those
obtained from the surface brightness equation:

\begin{equation}\label{eq-ang-diam}
\theta(\phi)=10^{0.2(S_V(\phi)-m_V(\phi))}
\end{equation}

where $\phi$ is the phase along the pulsational cycle and $\theta$ is
given in mas (see Fig.~\ref{fig-test2}). A glance at the
  figure shows that current solution shows a systematic difference  
with the interferometric measurements in the phase range between 0.2 and 0.45. 
This discrepancy is due to the limited accuracy of the calibration we are using 
to estimate the surface gravity. Typically, current calibration underestimates 
the gravity, in the above phase interval, and, in turn overestimates the radius.
The reader interested in a thorough analysis of this problem is referred to 
\citet{p78}. He suggested that the main culprit of the above discrepancy is the 
fact that we use grid of atmosphere models that neglect the microturbolence 
variation along the pulsation cycle. Furthermore, the problem in the $W$-band 
reflectivity, described above, also affects the surface gravity calibration. 
However, it is worth mentioning that a detailed analysis of the residuals between 
current estimates and interferometric measurements indicate a mean difference 
and a dispersion of the order of 3\%.

Assuming that the interferometric measurements are
minimally affected by systematic errors, we can use  this average difference to estimate 
our  systematic indetermination on $S_V$ through Eq.~\ref{eq-sb},
obtaining a value of $\sim 8\%$. Recalling that $S_V$ enters only in
the determination of the $\Delta$B, which in turn is only $\sim$ 5\%
of the radius, we found that the systematic uncertainty on our
estimate of $S_V$ implies a 0.4\% uncertainty on the radius.  

The agreement between the physical parameters obtained with our
procedure and those obtained independently from spectroscopy and
interferometry seems a valuable test of the accuracy in the derivation of
the surface brightness and of the $\Delta B$  term, thus supporting the
soundness of our approach.  
       
\section{Photometric and Spectroscopic data}
\label{sec-data}

After the validation of the quoted approach, we applied it to real data in order 
to obtain the radii of a sample of classical Galactic Cepheids. 

The photometric data and the radial velocities used in this work are
described in the following.

\subsection{Photometric data}
The original sample we have used in this work includes 175 Cepheids
observed in the Walraven VBLUW photometric system during 1962 and
1970--71 at the Leiden Southern Station (South Africa)
\citep{w64,p76}. It is $82\%$ complete for all Cepheids brighter than
$V=11.0$ magnitude at minimum of the light curve and with declination
lower than $+15\degr$. 

We have selected all the type I Cepheids (163 stars) included in this
sample, but we have retained only those stars for which reliable
radial velocity data were available in the literature (see next
section).
Furthermore, we have rejected also those stars having the colour--colour
loop at least in part outside the grid of models. This
behaviour can be observed for Cepheids having an extremely large loop
or a loop with a peculiar orientation in the colour--colour plane
(abnormal loop orientations can be caused by the presence of
companions or by peculiar chemical abundances).  
A complete list of the selected stars, with the photometric parameters
used in this work, is shown in Tab.~\ref{tab-stars}.  

\begin{table*}
\begin{center}
\caption{The names of Cepheids used in this work together with their
  period are listed in the first and second columns
  respectively. The third and fourth columns contain respectively the colour
excess $E(B-V)$ in the Johnson system, from \citet{lc07} (L) or those
from \citet{f95} (F) transformed in the Laney and Caldwell system by
using the relation from \citet{fa07}, and the $E(V-B)$ in the
Walraven system calculated from Eq. (\ref{eq-reddening}). In the last column
we list the source of the radial velocity data and, when more than one
source is taken into account the shift in radial velocity (km
s$^{-1}$), if any, is given in parentheses.}
\begin{tabular}{r @{} c @{} c @{} c @{} l}
\hline
\hline
Star &\hspace{0.5 cm} Period (days) &\hspace{0.5 cm} E(B--V) (Johnson)
&\hspace{0.3cm} E(V--B) (Walraven) &\hspace{0.3cm} Radial velocity\\
\hline
$\eta$ Aql&     7.1766   &  0.126 (L)  &	0.064  & \citet{bm87}\\
FM Aql &     6.1142    &  0.583 (L)  &	0.283  & \citet{gs98}	 \\     
FN Aql &     9.4822    &  0.491 (L)  &	0.214  & \citet{gs98}	 \\
TT Aql &      13.7546   &  0.417 (L)  &     0.217	 & \citet{b02}   \\
U Aql  &      7.0239   &  0.351 (L)  &     0.173	 & \citet{b02}   \\
V496 Aql &   6.8070    &  0.383 (L)  &     0.167	 & \citet{g81}    \\
RY CMa &     4.6782    &  0.254 (L)  &	0.110	& \citet{bm88}(+0.5), \citet{gs98}	   \\
SS CMa &     12.3620    &  0.590 (L)  &     0.258	& \citet{cc85}   \\
AQ Car &      9.7690   &   0.173 (L) &  0.07   & \citet{cc85} \\                    
ER Car &     7.7187     &  0.096 (F)  &  0.042  & \citet{le80}(+0.4), \citet{s55}  \\
l Car  &     35.5330    &  0.138 (L)  &	0.075  & \citet{ta97} \\	
U Car  &     38.7681    &  0.260 (L)  &     0.124	 & \citet{cc85}   \\
V Car    &   6.6967    &  0.170 (L)  &     0.075	 & \citet{s55}    \\
VY Car   &   18.9213    &  0.218 (L)  &     0.106	 & \citet{cc85}    \\
XX Car   &   15.7162   &  0.361 (L)  &     0.157	  & \citet{cc85}    \\
XY Car   &   12.4348   &  0.416 (L)  &     0.182	  &\citet{cc85}    \\
AZ Cen &     3.2107    &  0.159 (L)  &	0.068  & \citet{g81} \\                    
V Cen    &   5.4939    &  0.305 (L)  &     0.125	 & \citet{b02}    \\
V339 Cen &   9.4672     &  0.415 (L)  &     0.181	 & \citet{cc85}    \\
XX Cen   &   10.9558    &  0.267 (L)  &     0.116	  & \citet{cc85}    \\
AG Cru &      3.8373   &   0.171 (L) &  0.074  & \citet{g81}(+4.2), \citet{s55}\\                     
S Cru  &     4.6900    &  0.144 (L)  &     0.062	& \citet{g81}   \\
X Cru    &   6.2200    &  0.272 (F)  &     0.118	 & \citet{b02}     \\
W Gem    &   7.9141    &  0.253 (L)  &     0.123	 & \citet{i99}    \\
SV Mon &      15.2321   &  0.220 (L)  &     0.108 	 & \citet{i99}    \\
R Mus  &     7.5099    &  0.114 (F)  &	0.049  & \citet{le80}	 \\
UU Mus &     11.6364   &  0.414 (L)  &     0.180   & \citet{b02}    \\
S Nor  &     9.7549    &  0.180 (L)  &     0.078	& \citet{bb94}   \\
U Nor  &     12.6413   &  0.816 (L)  &     0.396	 & \citet{cc85}   \\
BF Oph &     4.0678    &  0.223 (L)  &	0.096  & \citet{g81} \\                    
Y Oph    &   17.1241   &  0.660 (L)  &     0.290	  &\citet{gs98}    \\
RS Ori &     7.5668    &  0.363 (L)  &	0.169  & \citet{gs98}, \citet{i99}(-0.3) \\
AP Pup &      5.0843 &  0.198 (F)  &  0.086  & \citet{s55} \\                    
AT Pup &      6.6650    &   0.207 (L) &	0.089  & \citet{g85} \\                    
RS Pup &     41.3876    &  0.454 (L)  &	0.197  & \citet{sc04}  \\
X Pup    &   25.9610    &  0.404 (L)  &     0.177	 & \citet{bm88}     \\
WX Pup   &   8.9382    &  0.303 (L)  &     0.138	 &\citet{s55}, \citet{bm88}(-1.7)\\
RV Sco &     6.0613    &  0.365 (L)  &	0.158	& \citet{le80}(+2.9), \citet{g81}, \citet{s55}(-1.1)\\
V482 Sco &   4.5279    &  0.334 (L)  &     0.145	 & \citet{g81}    \\
V500 Sco &   9.3166    &  0.618 (L)  &   0.263	 & \citet{s55}    \\
V636 Sco &   6.7966    &  0.207 (F)  &     0.089	 & \citet{s55}    \\
EV Sct &     3.0910    &  0.703 (L)  &	0.300  & \citet{mc93}(+0.4),
\citet{bb94}(+0.4), \citet{sc04} \\
SS Sct &      3.6712  &  0.324 (L)  &     0.140   & \citet{g81}   \\
X Sct    &   4.1981    &  0.589 (F)  &     0.257	  & \citet{mc93}    \\
Y Sct    &   10.3415  &  0.751 (L)  & 	0.364     & \citet{bm88}    \\
AP Sgr &      5.0579   &  0.179 (L)  &	0.077  & \citet{g81} \\                    
BB Sgr &     6.6370    &  0.302 (L)  & 	0.131  & \citet{g81} \\                    
U Sgr  &     6.7449   &  0.408 (L)  &     0.176	 & \citet{bb94}   \\
X Sgr    &   7.0122    &  0.284 (L)  &     0.122	  & \citet{wc89}    \\
Y Sgr    &   5.7733    &  0.187 (L)  &     0.089     & \citet{b02}\\
YZ Sgr   &   9.5534    &  0.286 (L)  &   0.127 &\citet{le80,bm88}\\
W Sgr    &   7.5947   &  0.103 (L)  &     0.048    & \citet{bb94}     \\
R TrA  &     3.3893   &  0.144 (L)  &	0.054 	& \citet{g81}	   \\
S TrA  &      6.3234   &  0.081 (L)  &     0.035	 &\citet{le80}(+1.4), \citet{g81}\\
AH Vel &      4.2271   &  0.070 (F)  & 	0.030  & \citet{g77} \\                    
AX Vel &     2.5929    &  0.239 (L)  &	0.103  & \citet{s55} \\                    
BG Vel &     6.9236    &  0.426 (F)  &	0.186  & \citet{s55} \\                    
RY Vel &     28.1270    &  0.545 (L)  &	0.240	& \citet{cc85}	   \\
\hline
\end{tabular}
\label{tab-stars}
\end{center}
\end{table*}

\begin{table*}
\begin{center}
\contcaption{}
\begin{tabular}{r @{} c @{} c @{} c @{} l}
\hline
\hline
Star &\hspace{1.0 cm} Period (days) &\hspace{1.0 cm} E(B--V) (Johnson)
&\hspace{1.0cm} E(V--B) (Walraven) &\hspace{1.0cm} Radial velocity\\
\hline
RZ Vel &     20.3969    &  0.294 (L)  &  0.129	& \citet{cc85}   \\
SX Vel &      9.5499   &  0.274 (L)  &     0.108	 & \citet{s55}   \\
SW Vel &      23.4744   &  0.346 (L)  &     0.153	 & \citet{cc85}   \\
T Vel  &      4.6397   &  0.289 (L)  &     0.122	 & \citet{s55}   \\
V Vel    &   4.3710   &  0.144 (L)  &     0.090	 & \citet{g85}    \\
\hline
\end{tabular}
\end{center}
\end{table*}

Before using the photometric data, we have corrected the colours for
the reddening effect following the procedure described in \citet{pl08}. In
particular we have obtained the colour excess $E(V-B)$ in the Walraven
system from: 
\begin{equation}\label{eq-reddening}
\frac{E(B-V)_J}{E(V-B)}=2.375-0.169(V-B)
\end{equation}   
where $(V-B)$ is the mean\footnote[3]{The mean is derived by applying
  the integral mean value theorem to the colour curve.} Walraven
colour of the considered star and 
$E(B-V)_J$ is the colour excess in the Johnson system. As for the
colour excess in the Johnson filters, we have referred to \citet{lc07}
or to \citet{f95} when the value was not available in the former
work. In the latter case, we have transformed the value of the colour excess from the 
Fernie's scale into the Laney \& Caldwell scale by using the relation recently 
provided by \citet{fa07}, namely $E(B-V)_{L\&C}=0.952E(B-V)_{Fernie}$ mag.
The $(U-W)$ colour excesses follow from the relation $E(U-W)=0.45E(V-B)$ \citep{pl08}. 

\subsection{Radial velocity data}
We have obtained the radial velocity data from different sources
available in the literature. The references of radial velocity data are
listed in Tab.~\ref{tab-stars} for each star of the sample. 

The main catalogues used in this work are
from: \citet{s55, g77, g81, g85, cc85, bm87, bm88, mc93, bb94, ta97, gs98,
  i99, b02, sc04}. The sample of radial velocity measurements was selected according to the 
following criteria. We neglected Cepheids with less than 10 measurements. 
For the Cepheids with multiple samples of radial velocity measurements,
we selected the data set characterized either by the most accurate measurements 
--typically better than $\sim$1 km/s-- and/or closer in time with photometric 
data. The latter criterion was adopted to minimize possible problems due to 
period changes and/or phase of maxima that are not very precise. In a few cases 
(nine stars), we have combined two or more data sets in order to improve the 
sampling along the radial velocity curve. To accomplish this goal, we corrected 
for possible shifts in radial velocity among the different data sets. The shift 
was determined by interpolating the most accurate sample of radial velocities 
and then by estimating the median distance --along the velocity axis-- between 
the fitted curve and the other radial velocity samples. The values of the shift 
we found are listed in Tab.~\ref{tab-stars}.

In order to obtain an accurate phasing between the photometric and
spectroscopic data, we recalculated the phases of radial velocity
measurements by using the same epoch and period as for the photometry.
The final sample used in this work consists of 63
Cepheids whose main characteristics are listed in Tab.~\ref{tab-stars}.

\section{Derivation of the radius of Cepheids}
\label{sec-radii}
\subsection{The projection factor $p$}
\label{sec-pfactor}

Using the photometric data and the radial velocities discussed in the
previous section, we are able to derive the mean radius for each star of our
sample following the procedure introduced in Secs. ~\ref{sec-cors} and 
  ~\ref{sec-cors-modif}. However, before proceeding we have to 
fix the value of the projection parameter $p$. 
In the recent literature there is a strong debate about the correct
value to use for this important parameter. Moreover, it is not
  clear yet if $p$ is constant over the entire period range covered by
  the Cepheids.
A detailed review of this subject can be found in \citet{b09} and there is no need to
repeat here the discussion outlined in that paper, but the
conclusion that can be drawn is that the $p$ value is uncertain at a
level of 5-10\%. The same uncertainties apply to the possible dependence of $p$ on the
Cepheid pulsation period. As a matter of fact, several authors find either a mild
\citep[see e.g.][]{nm07,ng09} or a strong \citep[see e.g.][]{g09}  
dependence on period.  On the contrary, \citet{g07}, using Cepheids with
HST parallaxes \citep{bm07} and interferometrically measured angular
diameters, find, convincingly, a value of $p=1.27\pm0.05$ with no
dependence on the period. The same result ($p=1.27\pm0.06$) was found
geometrically for $\delta$ Cephei by using interferometric
measurements, obtained with the CHARA Array, and published parallax \citep{mk05}.

Moreover, the projection factor $p$ is usually assumed to be constant 
along the pulsation cycle. However, this assumption was questioned by 
\citet{rb97}, since by using the changes of $p$ predicted by \citet{sk94}, 
they found that the radius determinations change by approximately  6\%. 
The change of the $p$ factor along the cycle has been the crossroad of several 
theoretical \citep{ms02} and empirical \citep{nf04} investigations, but we still lack 
firm quantitative estimates of the changes. This is the reason why it
is neglected in recent estimates of the Cepheid radii. We also neglected this 
change to perform comparison with similar data available in the 
literature.

Given this overall uncertainty, we decided to use two different values
of the $p$ factor: 1.36 and 1.27. These two values were chosen  because
they include the range of values found in the literature and, in particular
the value 1.36 was selected to allow comparison with our previous
works \citep{rb97, rr04}.

\subsection{Radius calculation}

The FORTRAN77 code used in this work performs a fit of the
magnitude curve, the $(V-B)$, $(U-W)$ colour curves and the radial
velocity curve using a Fourier fit, with a number of harmonics fixed
interactively by the user. Then it solves the CORS Eq. (\ref{eq-cors})
for the radius at an arbitrary phase both with and without the $\Delta
B$ term. The mean radius is, finally, calculated by integrating 
the radial velocity curve twice. 

Table ~\ref{tab-radii} gives the results of our procedure together
with the radii obtained by other authors. For three stars
  the CORS procedure does not converge and consequently the radius is
  not available. In particular we have listed the name of the star,
information on binarity, the period, the radius obtained with and
without the $\Delta B$ term and some results found in the literature
\citep{rr04, bs05, ra94}. In order to perform the comparison with similar estimates available in 
the literature, we rescaled their radius determinations according to our  
value of the projection factor. We found that our radius determinations  
agree quite well with those provided by \citet{ra94}, with a median difference 
of less than 1\% and a dispersion of 15\%. On the other hand, current 
radius evaluations are $\sim13\%$ smaller than those provided by \citet{rr04} 
and $\sim 21\%$ larger than those by \citet{bs05}, with a dispersion of the 
order of 14\% for both of them.

As further test, we have selected the Cepheids of our sample with
available interferometric measurements of the radius. In particular,
we compared the radii contained in \citet{kb04}, obtained with $p=1.36$, 
with our results corresponding to the same projection factor. The
stars in common with their sample are: X Sgr, $\eta$ Aql, W Sgr, Y
Oph, l Car. As shown in Tab.~\ref{tab-interf} our radii are completely
consistent with interferometric measurements, except for Y Oph, but the
error given by Kervella et al. is extremely large. 

\begin{table}
\begin{center}
\caption{ Interferometric radii derived by \citet{kb04} compared with
  the results of this work for $p=1.36$. In the first column there is the name of
  the Cepheids in common with the sample of Kervella et al, while in
  the second and third columns there are the interferometric radius
  and our estimation respectively.}  
\begin{tabular}{r @{} c @{} c}
\hline
\hline
Star &\hspace{0.5 cm} Kervella et al. &\hspace{0.5 cm} This work \\
\hline
X Sgr &\hspace{0.5 cm} $52.2^{+23}_{-12}$ &\hspace{0.5 cm} $52.4^{+3.7}_{-3.5}$\\
$\eta$ Aql &\hspace{0.5 cm} $59.3^{+5.3}_{-4.6}$ &\hspace{0.5 cm}
$59.0_{-4.2}^{+4.5}$\\
W Sgr &\hspace{0.5 cm} $56.4^{+30}_{-16}$ &\hspace{0.5 cm} $53.4^{+4.1}_{-3.8}$\\
Y Oph &\hspace{0.5 cm} $136^{+325}_{-56}$ &\hspace{0.5 cm} $90.7^{+6.9}_{-6.4}$\\
l Car &\hspace{0.5 cm} $191.2^{+7.6}_{-6.0}$ &\hspace{0.5 cm} $190.7^{+13.6}_{-12.7}$\\
\hline
\end{tabular}
\label{tab-interf}
\end{center}
\end{table} 

We found another estimation of the radius of l Car and $\eta$ Aql in
\citet{dj09} and \citet{j08} respectively. Using interferometric
observations made with the Sydney University Stellar Interferometer,
they found $R=169\pm 8 R_{\sun}$ for l Car and $R=39\pm 6 R_{\sun}$
for $\eta$ Aql, adopting a constant projection factor $p=1.30$. We 
rescaled their estimations using our projection factor and obtained
$R=177\pm 8 R_{\sun}$ for l Car, which is in agreement with our result
within 0.9 $\sigma$, while we are not in agreement with the radius of
$\eta$ Aql, $R=41\pm 6 R_{\sun}$.

\begin{table*}
\begin{center}
\caption{Radii derived in the present work compared with those
  previously determined by other authors \citep{rr04, bs05, ra94}. The
  third column gives information on binarity  according to the
  nomenclature from \citet{s03}: B=spectroscopic binary, 
  b=photometric companion with no sure physical relation,
  O=spectroscopic binary with known orbit, V=visual binary. The
  Cepheids with no binarity label are bona fide single stars.}
\begin{tabular}{r @{} c @{} c @{} c @{} c @{} c @{} c @{} c @{} c @{} c}
\hline
\hline
Star
&\hspace{0.2cm}Period(days)&\hspace{0.2cm}Binarity&\hspace{0.2cm}$\frac{R_{without\Delta
    B}^{p=1.36}}{R_{\sun}}$&\hspace{0.2cm}$\frac{R_{\Delta
    B}^{p=1.36}}{R_{\sun}}$&\hspace{0.2cm} $\frac{R_{without\Delta
    B}^{p=1.27}}{R_{\sun}}$&\hspace{0.2 cm}$\frac{R_{\Delta
    B}^{p=1.27}}{R_{\sun}}$&\hspace{0.2cm}
$\frac{R_{Ruoppo04}}{R_{\sun}}$&\hspace{0.2cm} 
$\frac{R_{Barnes05}}{R_{\sun}}$ &\hspace{0.2 cm} $\frac{R_{Rojo94}}{R_{\sun}}$\\ 
\hline
$\eta$ Aql&  7.1766   &  B   &    58.2       &59.0  &54.3 &55.1  &52.6  & 49.2  &58.1      \\
FM Aql &     6.1142    &  -   &    59.1       &58.8  &55.2 &54.9  &59.3  &  -    &  52.3 \\     
FN Aql &     9.4822    &  b   &    74.1       &75.5  &69.2 &70.5  &86.5  &   -   &  -  \\
TT Aql &      13.7546   &  -   &    103.6      &107.5 &96.7 &100.4  & -   &  -    &103.8 \\
U Aql  &      7.0239   &  O   &    48.8       &49.0  &45.6 &45.8  &47.0 &  -    & 53.1 \\
V496 Aql &   6.8070    &  O   &    47.2       &47.3  &44.1 &44.2  &61.2 &  -    &59.0 \\
RY CMa &     4.6782    &  B   &    30.1       &30.4  &28.1 &28.4  & -   &   -   &  42.2   \\
SS CMa &     12.3620    &  B   &     -	       &-     &- &-  &  -   &  -    & - \\
AQ Car &      9.7690   &  -   &    69.3       &72.3  &64.7 &67.5  &-     &  -    &   -   \\                    
ER Car &     7.7187     &  -   &    56.6       &58.2  &52.8 &54.3  &-     &   -   &  -  \\
l Car  &     35.5330    &  -   &    174.7      &190.7 &163.1 &178.1  &-    & 179.9 &  -    \\	
U Car  &     38.7681    &  B   &    -          &-     &- &-  & -   &  -    & - \\
V Car    &   6.6967    &  B   &    64.4       &68.1  &60.1 &63.6  &  -  &  -    &-  \\
VY Car   &   18.9213    &  B   &    124.9      &143.0 &116.6 &133.6  &  -  &112.2  &-      \\
XX Car   &   15.7162   &  B   &   85.7        &87.6  &80.0 &81.8  &  -  &  -    & - \\
XY Car   &   12.4348   &  -   &   83.2        &87.5  &77.7 &81.7  &  -  &  -    & - \\
AZ Cen &     3.2107    &  -   &    38.2       &39.2  &35.7 &36.6  &-     &  -    &  -  \\                    
V Cen    &   5.4939    &  -   &    53.1       &51.4  &49.6 &48.0  &  -  &39.6  &  -   \\
V339 Cen &   9.4672     &  V   &    68.0       &69.6  &63.5 &65.0  &  -  &  -    & - \\
XX Cen   &   10.9558    &  O   &   61.5        &63.6  &57.5 &59.4  &  -  & 67.2  & -    \\
AG Cru &      3.8373   &  B   &    29.1       &29.6  &27.2 &27.6  &-     &  -    &-    \\                     
S Cru  &     4.6900    &  -   &    46.5       &45.8  &43.5 &42.8  &  -  &  -    &-  \\
X Cru    &   6.2200    &  -   &   45.1        &45.6  &42.1 &42.6  &  -  &  -    &-  \\
W Gem    &   7.9141    &  -   &    71.3       &72.1  &66.5 &67.3  &86.4 &  -    & 58.8 \\
SV Mon &      15.2321   &  -   &    105.5      &114.5 &98.5 &106.9  & -   &  -    &115.4  \\
R Mus  &     7.5100    &  B   &    58.1       &58.2  &54.2 &54.4  & -   &  -    &  -  \\  
UU Mus &     11.6364   &  -   &    95.0      &94.7  &88.7 &88.4  & -   & 73.5  & -     \\
S Nor  &     9.7549    &  O   &    72.6       &74.2  &67.8 &69.3  & -    &73.8   &  -   \\
U Nor  &     12.6413   &  -   &    87.3       &82.1  &81.5 &76.6  & -   &75.0  & -    \\
BF Oph &     4.0678    &  B   &    46.8       &46.6  &43.7 &43.5  &44.8  & 30.1  &40.3     \\                    
Y Oph    &   17.1241   &  O   &   88.5        &90.7  &82.7 &84.7   &93.4 &92.1  & 98.3    \\
RS Ori &     7.5668    &  B   &    55.6       &56.4  &51.9 &52.7  & -   &   -   &  -  \\
AP Pup &      5.0843 &  B   &    45.3       &45.9  &42.3 &42.8  &-     &  -    &-      \\                    
AT Pup &      6.6650    &  B   &    50.0       &49.3  &46.7 &46.0  &-     &  -    &-   \\                    
RS Pup &     41.3876    &  -   &    223.5      &228.8 &208.7 &213.7  & -   & 205.1     &  -    \\
WX Pup   &   8.9382    &  -   &    64.5       &64.7  &60.2 &60.4  &  -  &  -    &56.8 \\
X Pup    &   25.9610    &  -   &   134.5       &146.7 &125.6 &137.0  &  -  &  -    &118.5  \\
RV Sco &     6.0613    &  B   &    53.9       &53.4  &50.3 &49.9  & -   &  -    &-    \\
V482 Sco &   4.5279    &  B   &    52.1       &52.8  &48.6 &49.3  &  -  &  -    &- \\
V500 Sco &   9.3166    &  b   &    72.5       &71.1  &67.7 &66.4  &  -  &  -    & - \\
V636 Sco &   6.7966    &  O   &    65.5       &66.4  &61.2 &62.0  &  -  &  -    & - \\
EV Sct &     3.0910    &  B:  &    94.9       & -    &88.6 &-  & -       &  36.1   &  -  \\
SS Sct &      3.6712  &  -   &    39.2       &38.9  &36.6 &36.3  &44.2 &  -    & 36.0 \\
X Sct    &   4.1981    &  b   &   58.4        &53.5  &54.5 &50.0  &  -  &  -    & - \\
Y Sct    &   10.3415  &  b   &   65.4        &76.6  &61.0 &71.5  &  -  &  -    & 84.3 \\
AP Sgr &      5.0579   &  B   &    48.6       &47.9  &45.4 &44.7  &62.3  &  -    & 46.1   \\                    
BB Sgr &     6.6370    &  B   &    44.2       &43.8  &41.3 &40.9  &61.3  & 45.6  & 54.1  \\                    
U Sgr  &     6.7449   &  B   &    64.6       &65.7  &60.4 &61.3  &82.5 &49.1   &55.6    \\
W Sgr    &   7.5947   &  O   &    53.2       &53.4  &49.7 &49.9  &62.6 &  -    & 50.8 \\
X Sgr    &   7.0122    &  O   &   54.3        &52.4  &50.7 &48.9  & 59.8&  -    &44.1 \\
Y Sgr    &   5.7733    &  B   &   55.7        &55.2  &52.0 &51.5  &56.2 &  -    & 48.3 \\
YZ Sgr   &   9.5534    &  B   &   66.8        &69.3  &62.4 &64.7  &95.8 &  -    & 69.4 \\
R TrA  &     3.3893   &  B   &    35.0       &35.1  &32.7 &32.8  & -   &  -    &  -  \\
S TrA  &      6.3234   &  -   &    54.1       &54.8  &50.5 &51.2  & -   &  -    & - \\
AH Vel &      4.2271   &  B   &    55.9       &51.7  &52.2 &48.3  &-     &  -    &-    \\                    
AX Vel &     2.5928    &  -   &    53.8       &54.1  &50.2 &50.5  &-     &  -    &  -  \\                    
BG Vel &     6.9236    &  b   &    66.0       &69.7  &61.6 &65.1  &-     &   -   &  -  \\                    
RY Vel &     28.1270    &  -   &    172.3      &181.2 &160.9 &169.2  & -   & 136.4 &  -    \\
\hline
\end{tabular}
\label{tab-radii}
\end{center}
\end{table*}

\begin{table*}
\contcaption{}
\begin{center}
\begin{tabular}{r @{} c @{} c @{} c @{} c @{} c @{} c @{} c @{} c @{} c}
\hline
\hline
Star
&\hspace{0.2cm}Period(days)&\hspace{0.2cm}Binarity&\hspace{0.2cm}$\frac{R_{without\Delta
   B}^{p=1.36}}{R_{\sun}}$&\hspace{0.2cm}$\frac{R_{\Delta
   B}^{p=1.36}}{R_{\sun}}$&\hspace{0.2cm} $\frac{R_{without\Delta
   B}^{p=1.27}}{R_{\sun}}$&\hspace{0.2 cm}$\frac{R_{\Delta
   B}^{p=1.27}}{R_{\sun}}$&\hspace{0.2cm}
$\frac{R_{Ruoppo04}}{R_{\sun}}$&\hspace{0.2cm} 
$\frac{R_{Barnes05}}{R_{\sun}}$ &\hspace{0.2 cm} $\frac{R_{Rojo94}}{R_{\sun}}$\\ 
\hline
RZ Vel &     20.3969    &  -   &    130.8      &137.0 &122.1 &128.0  & -   & 110.6 &  -   \\
SX Vel &      9.5499   &  -   &    64.3       &65.1  &60.1 &60.8  & -   &  -    & - \\
SW Vel &      23.4744   &  -   &    140.1      &151.9 &130.3 &141.8  & -   &122.2 &  -    \\
T Vel  &      4.6397   &  B   &    57.2       &57.3  &53.4 &53.5  & -   &40.0   &  -   \\
V Vel    &   4.3710   &  B   &    34.4       &34.2  &32.1 &31.9  &  -  &  -    & - \\
\hline
\end{tabular}
\end{center}
\end{table*}

\section{The Period--Radius relation}

\label{sec-pr}
Since the radii obtained in this work are consistent with those
available in the literature, we have investigated the Period--Radius
relation for the present data set.

\begin{figure}
\includegraphics[width=84mm]{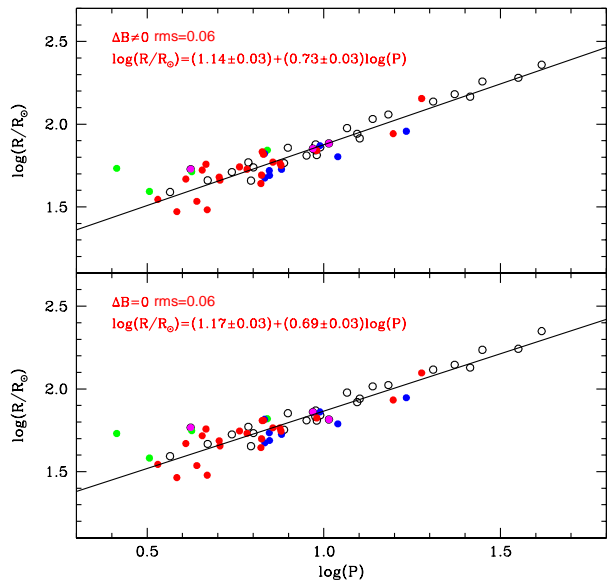}
\caption{Period--Radius relations obtained in the case with $\Delta B$
  term (bottom panel) and without $\Delta B$ term (upper panel), using
  all the Cepheids of our sample, with the exception of first overtone
  pulsators (green points). Single stars and visual binaries are
  plotted as black circles, spectroscopic binaries as red points,
  spectroscopic binaries with known orbit as blue points and binaries
  with photometric companion with unknown physical relations as magenta
  points.} 
\label{fig-pr-all}
\end{figure}

In order to derive the Period--Radius relation, we neglected the first overtone pulsators, 
since these objects have, at fixed period, systematically larger radii than 
fundamental pulsators. To pin point first overtone pulsators in our sample we followed 
the classification given by \citet{f95}. Data plotted in Fig. 8 show the 
radii for the entire Cepheid sample. The different colors mark Cepheids 
in different types of binary systems. Following the classification suggested 
by \citet{s03}, the black circles display single stars and visual binaries (V), 
the red points show the spectroscopic binaries (B), the blue points the 
spectroscopic binaries with known orbits (O) and the magenta points mark the 
binaries with photometric companions, but their binarity has to be confirmed (b), 
while the green circles show the position of the first overtone pulsators. 
We ended up with a sample of 56 Cepheids and they have been used to derive a new 
Period–-Radius relation for Galactic Cepheids. We performed a linear fit in the 
form $\log R = a\log P + b$, with the radius in solar units. Finally, by
using $p=1.36$, we have found the following relations:
\begin{equation}
log R = (0.69\pm 0.03)\log P + (1.17\pm 0.03) 
\label{eq-pr-all-db0}
\end{equation}
\begin{equation}
\log R = (0.73\pm 0.03)\log P + (1.14\pm 0.03) 
\label{eq-pr-all-db}
\end{equation}
in the case without and with the $\Delta B$ term, respectively.
The standard deviation of the above Period--Radius relations is 0.06 dex. A glance
at the data plotted in Fig.~\ref{fig-pr-all} indicates that a significant fraction of the 
scatter is due to binaries. The flux of the companions affect the radius 
estimates. This effect is more significant in optical and in ultraviolet 
bands than in near infrared bands, because the Cepheid companions are 
typically blue stars \citep[see e.g.][]{s03}. In order to provide  
a tighter Period--Radius relation, we neglected all the spectroscopic binaries 
from the fit (types B and O of the above classification) and the first 
overtone candidate X Sct (see Sec.~\ref{sec-xsct}). The final sample reduces 
to 26 Cepheids and the corresponding Period–-Radius relations for 
$p=1.36$ are the following:
\begin{equation}
\log R=(0.71\pm 0.03)\log P + (1.16\pm 0.03)
\label{eq-pr-xsct-db0}
\end{equation} 
\begin{equation}
\log R=(0.75\pm 0.03) \log P + (1.13\pm 0.03) 
\label{eq-pr-xsct-db}
\end{equation} 
with an rms of 0.04 and 0.03 dex in the case without and with the 
$\Delta B$ term, respectively.
The exclusion of the binary stars does not affect significantly the 
coefficients of the Period–-Radius relation, but it significantly 
decreases the scatter of the linear fit.

The fitted equations for the case $p=1.27$ are the following:
\begin{equation}
\log R=(0.71\pm0.03)\log P+(1.13\pm0.03)
\label{eq-pr-newp-db0} 
\end{equation}  
\begin{equation}
\log R=(0.75\pm0.03)\log P+(1.10\pm0.03)
\label{eq-pr-newp-db} 
\end{equation}  
for the case without and with the $\Delta B$ term respectively. The
rms errors of these relation are the same as for the case with $p$=1.36.

\begin{figure}
\includegraphics[width=84mm]{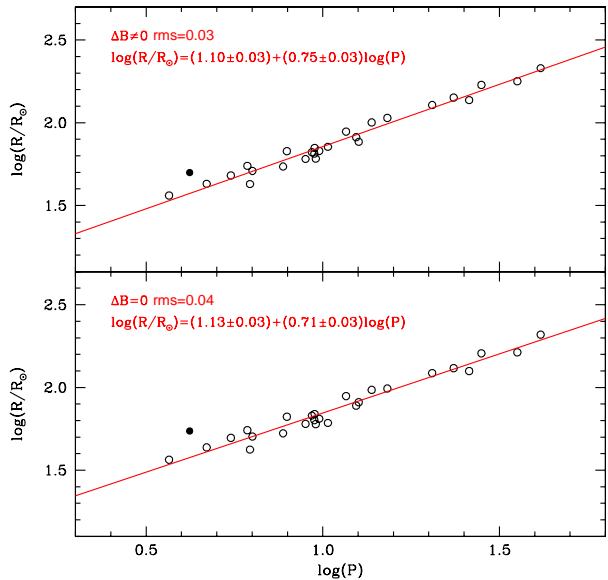}
\caption{Period--Radius relations for the final sample of Cepheids in
  the case with $\Delta B$ (top panel) and without $\Delta B$ (bottom
  panel). In each panel the fitting equation and the scatter around
  the fit are also shown. The possible first overtone X Sct is also
  plotted (filled circle) but it is excluded from the fit.}    
\label{fig-pr}
\end{figure}

In order to validate current Period-Radius relations, we performed a 
detailed comparison with similar theoretical and empirical relations    
available in the literature. The zero-points and the slopes are listed 
in Tab.~\ref{tab-other-pr}. 

In order to compare theory and observations we selected three 
different predictions concerning the Period--Radius relations at 
solar chemical composition. In particular the Period--Radius relation 
provided by \citet{ab99} by using linear, nonadiabatic, 
convective models constructed assuming a canonical (i.e. based 
on evolutionary models that neglect convective core overshooting 
during central hydrogen burning phase, rotation and mass loss) 
Mass--Luminosity relation. Moreover, we also adopted predictions
provided by \citet{bc98} and by \citet{pb03} by using,
nonlinear, convective models constructed using both canonical and
noncanonical Mass--Luminosity relations (mimicking the increase in 
luminosity caused by overshooting). To provide a robust comparison between
theory and observations, we performed a new estimate of the Period--Radius 
relations using only the models of the last two data sets covering 
the same period range as of the current sample of Galactic Cepheids. 
The zero-points and the slopes listed in the first three entries of 
Tab.~\ref{tab-other-pr} indicate that they agree quite well with empirical Period--Radius 
relations obtained by neglecting the $\Delta B$ term. Theoretical and 
empirical uncertainty do not allow us to constrain the precision of 
the Period--Radius relations based on the two different projection 
factors.

\begin{figure}
\includegraphics[width=84mm]{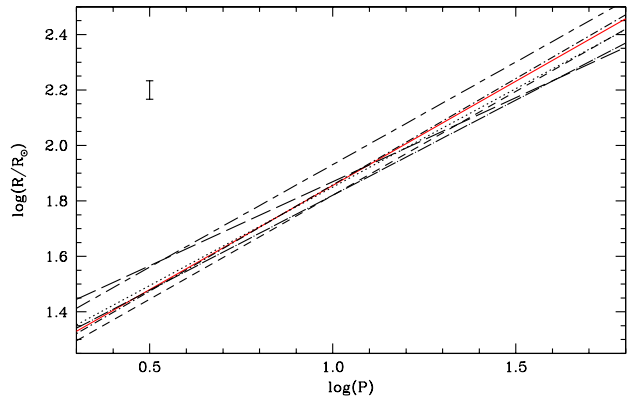}
\caption{Our best fit Period--Radius relation (red solid line) is
  plotted together with those obtained by \citet{ab99} (theoric
  relation, dot line),
  \citet{gf98} (short dashed line), \citet{rr04} (short dashed long dashed
  line), \citet{rb97} (long dashed line), \citet{kb04} (dot short dashed line)
  and \citet{g07} (dot long dashed line). The error bar represents the
standard deviation of the current relation (0.03 dex).}    
\label{fig-pr-other}
\end{figure}  

\begin{table*}
\begin{center}
\caption{Coefficients of the Period--Radius relation from the
  literature and of our work. The first two columns contain the slope and
  the intercept of the linear relation, the third column contains the
  reference, the fourth column the technique used to derive the
  Period--Radius relation and the last one the number of stars used in
  the fit.}
\begin{tabular}{l @{} l @{} l @{} l @{} l}
\hline
\hline
$a$ &\hspace{0.3 cm} $b$ &\hspace{0.3cm} $Source$ &\hspace{0.3cm}
$Method$ &\hspace{0.3cm}$N$\\
\hline
$0.71$ &\hspace{0.3 cm} $1.14$ &\hspace{0.3cm} \citet{ab99}$^{a}$
&\hspace{0.3cm} Theory (Canonical models; Solar metallicity) &\hspace{0.3cm} -\\
$0.673\pm 0.009$ &\hspace{0.3 cm} $1.164\pm 0.010$ &\hspace{0.3cm} \citet{bc98}
&\hspace{0.3cm} Theory (Canonical+Noncanonical models; Solar
metallicity)&\hspace{0.3cm} -\\
$0.699\pm 0.011$ &\hspace{0.3 cm} $1.150\pm 0.012$ &\hspace{0.3cm} \citet{pb03}
&\hspace{0.3cm} Theory (Canonical+Noncanonical models; Solar metallicity)&\hspace{0.3cm} -\\
$0.75\pm0.02$ &\hspace{0.3 cm} $1.07\pm0.02$ &\hspace{0.3cm} \citet{gf98}$^{b}$
&\hspace{0.3cm} Surf. Brightness (variable p)&\hspace{0.3cm} 28\\
$0.74\pm0.03$ &\hspace{0.3 cm} $1.12\pm0.03$ &\hspace{0.3cm} \citet{ra94}
&\hspace{0.3cm} Surf. Brightness ($p=1.31$)&\hspace{0.3cm} 54\\
$0.606\pm0.037$ &\hspace{0.3 cm} $1.263\pm0.033$ &\hspace{0.3cm} \citet{rb97}
&\hspace{0.3cm} Surf. Brightness ($\Delta B\ne 0$, $p=1.36$)&\hspace{0.3cm} 64\\
$0.767\pm0.009$&\hspace{0.3 cm} $1.091\pm0.011$&\hspace{0.3cm}
\citet{kb04}&\hspace{0.3cm} Interferometry ($p=1.36$)&\hspace{0.3cm} 8\\
$0.686\pm0.036$&\hspace{0.3 cm} $1.134\pm0.034$&\hspace{0.3cm}
\citet{g07}&\hspace{0.3cm} Interferometry ($p=1.27$)&\hspace{0.3cm} 5\\
$0.747\pm0.028$ &\hspace{0.3 cm} $1.071\pm0.025$ &\hspace{0.3cm} \citet{tb02}
&\hspace{0.3cm} Modified Baade--Wesselink ($p=1.31$)&\hspace{0.3cm} 13\\
$0.69\pm0.09$ &\hspace{0.3 cm} $1.18\pm0.08$ &\hspace{0.3cm} \citet{rr04}
&\hspace{0.3cm} New CORS ($\Delta B= 0$, $p=1.36$)&\hspace{0.3cm} 20\\
$0.74\pm0.11$ &\hspace{0.3 cm} $1.19\pm0.09$ &\hspace{0.3cm} \citet{rr04}
&\hspace{0.3cm} New CORS ($\Delta B\ne 0$, $p=1.36$)&\hspace{0.3cm} 16\\
$0.71\pm0.03$ &\hspace{0.3 cm} $1.16\pm0.03$ &\hspace{0.3cm} This Work
&\hspace{0.3cm} New CORS ($\Delta B=0$, $p=1.36$ )&\hspace{0.3cm} 26\\
$0.75\pm0.03$ &\hspace{0.3 cm} $1.13\pm0.03$ &\hspace{0.3cm} This Work
&\hspace{0.3cm} New CORS ($\Delta B\ne0$, $p=1.36$)&\hspace{0.3cm} ``\\
$0.71\pm0.03$ &\hspace{0.3 cm} $1.13\pm0.03$ &\hspace{0.3cm} This Work
&\hspace{0.3cm} New CORS ($\Delta B=0$, $p=1.27$ )&\hspace{0.3cm} ``\\
$0.75\pm0.03$ &\hspace{0.3 cm} $1.10\pm0.03$ &\hspace{0.3cm} This Work
&\hspace{0.3cm} New CORS ($\Delta B\ne0$, $p=1.27$)&\hspace{0.3cm} ``\\
\hline
\end{tabular}
\label{tab-other-pr}
\end{center}
\begin{flushleft}
$^{a}$\footnotesize{Errors on the coefficients are not indicated in
    the original paper (\citet{ab99}, Tab.5).}\\
$^{b}$\footnotesize{In order to have uniformity in the comparison with other
relations, the listed uncertainties are the standard deviations on 
the parameters, which have been recalculated by fitting their data.}
\end{flushleft}
\end{table*}

The values given in Tab.~\ref{tab-other-pr} indicate that current Period--Radius 
relations agree, within the errors, with similar empirical estimates.   
On the other hand, the Period--Radius relation provided by
\citet{rb97}, using the same method, gives a smaller zero-point, but  
a steeper slope.

It is interesting to note that the zero--points of our relations seem
to change in opposite directions by including or excluding the $\Delta
B$ term with respect to those by \citet{rr04}. The main result
derived from Tab.~\ref{tab-other-pr} is that the agreement of the
zero--point of our relations with those of other authors (excluding
the theoretical result) improves by
including the $\Delta B$ term and by decreasing the projection factor
from 1.36 to 1.27. Therefore, we will consider the equation:
\begin{equation}\label{eq-best-pr}
\log R=(0.75\pm0.03)\log P+(1.10\pm0.03)
\end{equation}
as our best fit Period--Radius relation. A plot of our relations for
$p=1.27$ is shown  in Fig.~\ref{fig-pr}; the plot for $p=1.36$ is not
shown because the points are only shifted vertically by the same
offset of 0.03. The projection factor does not affect the slope of the
Period--Radius relation only when it is assumed to be constant over
the period range. A visual comparison of Eq.~\ref{eq-best-pr} with other
relations is shown in Fig.~\ref{fig-pr-other}.

\subsection{X Sct: a possible first overtone pulsator}\label{sec-xsct}
To test  the first overtone nature of X Sct, we
estimated the radius predicted from our Period--Radius relations for a
Cepheid with the same period: $R=36.9^{+2.7}_{-2.4}$, for $p=1.27$ and
$R=39.6^{+2.8}_{-2.6}$ with $p=1.36$, in units solar radii. The values
obtained from the CORS method are $R=50.0$ and
$R=53.5$, for $p=1.27$ and $p=1.36$
respectively (see Tab.~\ref{tab-radii}). They differ by more than
$4\sigma$ from the predicted values, being larger, as expected for a
first overtone pulsator.   

As a further test we have also estimated the radius predicted by
the theoretical Period--Radius relation for first overtone Cepheids
obtained by \citet{bgm01}:
\begin{eqnarray}
\log R=(0.755\pm0.007)\log P+(1.250\pm 0.005)\\
\log R=(0.737\pm0.005)\log P+(1.219\pm 0.004)
\end{eqnarray}
for canonical and non-canonical models respectively. The radius 
corresponding to the period of X Sct is $R=52.5^{+1.7}_{-1.1}$ and
$R=47.7^{+0.7}_{-0.6}$ using the canonical and the non-canonical relation
respectively. It is evident that the value obtained in the case of the
canonical relation is in excellent agreement with that obtained from
CORS with $p=1.36$. The value obtained from non-canonical
relation is consistent only within  $\sim 3.3\sigma$ with the CORS value
for $p=1.27$.

Furthermore, we have also compared the radii of X Sct with those
of the first overtone Cepheid AH Vel ($P=4.23$) and of the
fundamental pulsator V Vel ($P=4.37$), obtained using CORS (see
Tab.~\ref{tab-radii}). As a 
result we find that the radius of X Sct is much more consistent with a
first overtone pulsation.

On the basis of this test, we are confident that X Sct can be a good first
overtone candidate, but further analysis is required to confirm this
hypothesis. 

\section{The Period--Luminosity relation}
\label{sec-pl}
The estimated radii and effective temperatures are used to derive the
intrinsic luminosities of the investigated Cepheids and, in turn their
absolute visual magnitude.
To this aim we assumed the solar visual magnitude in the V band to
be -26.75 \citep{h85}, which gives an absolute visual magnitude of
+4.82 and an absolute bolometric magnitude of +4.75, with a bolometric
correction of -0.07. Furthermore, we calculated the bolometric
correction from Eq.~\ref{eq-bc}.
In order to evaluate the absolute magnitude uncertainty, we
assumed the scatter around the fitted $\log T_{eff}(V-B, U-W)$ and
$BC(V-B, U-W)$ surfaces as the uncertainty of the effective
temperature and bolometric correction respectively. The values of
absolute magnitudes and their uncertainties, $\delta M_V$, are listed
in Tab.~\ref{tab-mv} for the case $p=1.27$.  

The Period--Luminosity relation is obtained by a weighted linear fit of
the relation $M_V=a\log P+b$ and the resulting coefficients are listed
in Tab.~\ref{tab-pl-coeff} together with some results from other
authors. We list the coefficients obtained using both  values of
the projection factor, $p=1.27$ and $p=1.36$. 

As for the Period--Radius relation, the projection factor influences
only the zero--point of the Period--Luminosity relation, leaving the
slope unchanged.

From Tab.~\ref{tab-pl-coeff} it is evident that the slope of our
Period--Luminosity relation is steeper than those of the relations by
\citet{fa07} and \citet{bj03}, although they are consistent within the
errors. On the other hand, it is in excellent agreement with the slope of the
relations by \citet{gf98} and \citet{kb04}. 

The agreement of the zero-point with other works is better if
we choose the projection factor $p=1.27$, while in 
the case with $p=1.36$ the Period--Luminosity relation seems to make
Cepheids too luminous, particularly if we compare our relation with that
found by \citet{fa07} or \citet{gf98}. We note that the systematic
errors on the zero--point of the Period--Luminosity relations obtained
for the two values of p is $\sim 0.15$ mag, i.e. comparable with the
statistical error on the zero--point itself.

\begin{figure}
\includegraphics[width=84mm]{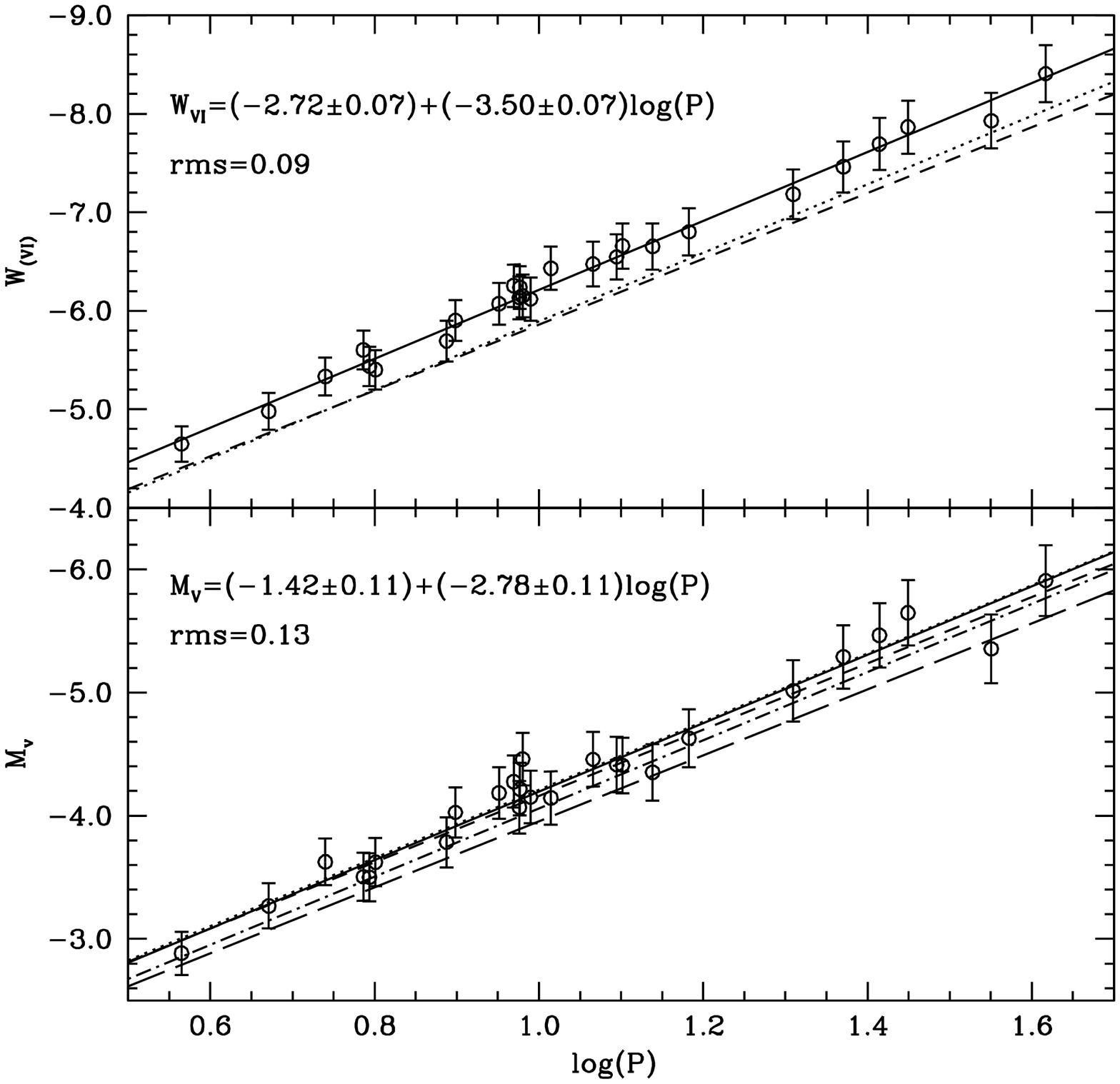}
\caption{Bottom panel: our Period--Luminosity relation ($p=1.27$) (red
  solid line) is plotted together with those obtained by \citet{kb04} (dotted line),
  \citet{bj03} (short dashed line), \citet{fa07} (long dashed
  line) and \citet{gf98} (dot short dashed line). Upper panel: the
  Period--Wesenheit relation obtained using 
  the projection factor $p=1.27$ (red solid line). The other two
  plotted fits are from \citet{fa07} (dotted line) and \citet{bm07}
  (short dashed line). 
  The rms around the fitted relations are indicated in both panels.}    
\label{fig-pl}
\end{figure}

A plot showing the comparison of our Period--Luminosity relation ($p=1.27$)
with those from \citet{bj03}, \citet{kb04}, \citet{fa07} and
\citet{gf98} is represented in Fig.~\ref{fig-pl}.  

\begin{table*}
\begin{center}
\caption{The data used to fit the Period--Luminosity and Period--Wesenheit
  relations are listed for the Cepheids of our sample. The absolute
  magnitudes $M_V$ used to derive the Wesenheit magnitudes are
  calculated as described in the text, while the
  $(\left \langle V \right \rangle- \left \langle I_C \right \rangle)$ colour
  in the last column is from \citet{g99}. For V500 Sco, W Gem, WX Pup,
  X Cru and Y Sct the  $(\left \langle V \right \rangle-\left \langle I_C \right \rangle)$ colour is not available in
  \citet{g99} and we have used the value from \citet{cc87}.} 
\begin{tabular}{r @{} c @{} c @{} c @{} c @{} c @{} c @{} c}
\hline
\hline
Star name &\hspace{0.5 cm} Period (days) &\hspace{0.5 cm} $T_{eff}$ (K) &\hspace{0.5 cm} $M_V$ (mag) 
&\hspace{0.5cm} $\delta M_V$ (mag) &\hspace{0.5cm} $W_{VI_C}$ (mag) &\hspace{0.5cm}
$\delta W_{VI_C}$ (mag) &\hspace{0.5cm} $(\left \langle V \right \rangle-\left \langle I_C \right \rangle)$ (mag)\\
\hline
FM Aql      &  6.11423    &5746   & -3.50  & 0.17    &  -5.60   & 0.18  & 1.531  \\
FN Aql      &  9.48224    &5731   & -4.22  & 0.18    &  -6.24   & 0.19  & 1.389  \\
TT Aql      &  13.7546    &5293   & -4.35  & 0.19    &  -6.65   & 0.19  & 1.409  \\
AQ Car      &  9.76896    &5627   & -4.15  & 0.18    &  -6.12   & 0.19  & 0.982  \\
ER Car      &  7.7187     &5642   & -3.78  & 0.20    &  -5.69   & 0.21  & 0.865  \\
l Car       &  35.5330    &4843   & -5.35  & 0.22    &  -7.93   & 0.22  & 1.177  \\
XY Car      &  12.43483   &5503   & -4.41  & 0.19    &  -6.55   & 0.19  & 1.342  \\
V Cen       & 5.49392     &6069   & -3.62  & 0.17    &  -5.33   &0.17& 1.041\\
V339 Cen    & 9.4672      &5591   & -4.07  & 0.18    &  -6.13   & 0.19  & 1.314  \\
S Cru       & 4.68997     &5956   & -3.27  & 0.17    &  -4.98   & 0.17  & 0.846  \\
X Cru       &6.219970     &5716   & -3.50  & 0.19    &  -5.44   & 0.20  & 1.090\\
W Gem       & 7.914130    &5854   & -4.03  & 0.18    &  -5.90   & 0.18  & 1.043\\
SV Mon      &  15.2321    &5391   & -4.63  & 0.19    &  -6.80   & 0.20  & 1.119  \\
UU Mus      & 11.63641    &5648   & -4.46  & 0.19    &  -6.47   & 0.19  & 1.294  \\
U Nor       & 12.64133    &5476   & -4.41  & 0.19    &  -6.66   & 0.19  & 1.874  \\
RS Pup      & 41.3876     &5087   & -5.91  & 0.22    &  -8.41   & 0.23  & 1.531  \\
X Pup       & 25.9610     &5368   & -5.46  & 0.21    &  -7.69   & 0.21  & 1.365  \\
WX Pup      &8.938250     &5809   & -4.18  & 0.18    &  -6.07   & 0.18  & 1.108\\
V500 Sco    & 9.316650    &5828   & -4.28  & 0.18    &  -6.25   & 0.19  & 1.527\\ 
SS Sct      &   3.671253  &5970   & -2.88  & 0.16    &  -4.65   & 0.17  & 1.086  \\
Y Sct       & 10.341504   &5523   & -4.14  & 0.18    &  -6.43   & 0.19  & 1.810\\
S TrA       &  6.32344    &5810   & -3.62  & 0.17    &  -5.40   & 0.18  & 0.796  \\
RY Vel      & 28.1270     &5411   & -5.65  & 0.21    &  -7.86   & 0.21  & 1.532  \\
RZ Vel      & 20.3969     &5316   & -5.01  & 0.20    &  -7.18   & 0.21  & 1.208  \\
SX Vel      &  9.54993    &5992   & -4.46  & 0.21    &  -6.15   & 0.22  & 0.996  \\
SW Vel      &  23.4744    &5355   & -5.29  & 0.21    &  -7.46   & 0.21  & 1.272  \\
\hline
\end{tabular}\label{tab-mv}
\end{center}
\end{table*}

\begin{table*}
\begin{center}
\caption{Coefficients of the Period--Luminosity and of the
  Period--Wesenheit relations obtained in this work and from other authors.} 
\begin{tabular}{c @{} c @{} c @{} c @{} c}
\hline
\hline
Band &$a$ &\hspace{1 cm} $b$ &\hspace{1 cm} rms   &\hspace{1 cm} $Source$ \\
\hline
$V$&\hspace{1cm} $-2.77\pm0.07$ &\hspace{1cm} $-1.29\pm0.08$ &\hspace{1cm} 0.20
&\hspace{1cm} \citep{gf98}\\
$V$&\hspace{1cm} $-2.77\pm0.07$ &\hspace{1cm} $-1.44\pm0.10$ &\hspace{1cm} -
&\hspace{1cm} \citep{kb04}\\
$V$&\hspace{1cm} $-2.68\pm0.08$ &\hspace{1cm} $-1.27\pm0.02$ &\hspace{1cm} 0.17
&\hspace{1cm} \citep{fa07}\\
$V$&\hspace{1cm} $-2.69\pm0.17$ &\hspace{1cm} $-1.47\pm0.28$ &\hspace{1cm} 0.19
&\hspace{1cm} \citep{bj03}\\
$V$&\hspace{1cm} $-2.78\pm0.11$ &\hspace{1cm} $-1.57\pm0.11$ &\hspace{1cm} 0.13
&\hspace{1cm} This Work (p=1.36)\\
$V$&\hspace{1cm} $-2.78\pm0.11$ &\hspace{1cm} $-1.42\pm0.11$ &\hspace{1cm} 0.13
&\hspace{1cm} This Work (p=1.27)\\
$W_{VI_C}$&\hspace{1cm}$-3.48\pm0.07$ &\hspace{1cm}
$-2.41\pm0.02$&\hspace{1cm} 0.17 &\hspace{1cm} \citep{fa07}\\
$W_{VI_C}$&\hspace{1cm}$-3.34\pm0.17$ &\hspace{1cm}
$-2.52\pm0.17$&\hspace{1cm} 0.09 &\hspace{1cm} \citep{bm07}\\
$W_{VI_C}$&\hspace{1cm}$-3.50\pm0.07$ &\hspace{1cm}
$-2.87\pm0.07$&\hspace{1cm} 0.09 &\hspace{1cm} This Work (p=1.36)\\
$W_{VI_C}$&\hspace{1cm}$-3.50\pm0.07$ &\hspace{1cm}
$-2.72\pm0.07$&\hspace{1cm} 0.09 &\hspace{1cm} This Work (p=1.27)\\
\hline
\end{tabular}
\label{tab-pl-coeff}
\end{center}
\end{table*}

Using the V band absolute magnitudes and the intensity--mean
$(\left \langle V \right \rangle-\left \langle I_C \right \rangle)$ colour tabulated in \citet{g99} for all our Cepheids,
except for V500 Sco, W Gem, WX Pup, X Cru and Y Sct, for which we 
used the values from \citet{cc87}, we derived the reddening free
Wesenheit magnitude $W_{VI_C}= M_V - R_{VI_C}(\left \langle V \right \rangle-\left \langle I_C \right \rangle)$, where the
colour coefficient $R_{VI_C}=2.55$ is adopted from \citet{cm00} (see
Tab.~\ref{tab-mv} for $W_{VI_C}$ values and their uncertainty $\delta W_{VI_C}$). 
The Period--Wesenheit relation, $W_{VI_C}=a\log P + b$, has been
derived through a weighted linear fit using both  values of the
projection factor; the results are listed in Tab.~\ref{tab-pl-coeff}
together with those obtained from \citet{fa07} and \citet{bm07}. 

The slope of our Wesenheit relation is in agreement with
the results from other authors, using both values of the projection factor.
The variation of the projection factor influences the zero--point,
which, for $p=1.36$ seems to be too luminous, up to 0.46 mag if we
consider the one by \citet{fa07}. This result, obviously, reflects the
behaviour of the obtained Period--Luminosity relation (see above). For
$p=1.27$, the Wesenheit zero--point results to be brighter by 0.20 mag
than the one by \citet{bm07} and by 0.31 mag than the one by \citet{fa07}.

The implications of these results for the extragalactic distance scale
calibration will be addressed in a forthcoming paper (Molinaro et
al. in preparation).

\section{Conclusions}
\label{sec-conclusion}
In this work we have developed a modified version of the CORS method
in the Walraven filters, following the approach already adopted by  
\citet{rr04} in the Str\"omgren system, to derive the radii of a sample
of 63 Galactic Cepheids observed in the Southern hemisphere. The
adopted photometric data are the results of an extensive observational 
program \citep{w64, p76, pl07}, whereas for the radial velocities we have
used data available in the literature from various authors.

The observations were performed in the Walraven five bands system (V,
B, L, U, W) well-suited to measure spectral characteristics of 
stars like Cepheids and RR Lyrae. 
Even though the Walraven photometric system is not available in modern
observational facilities, the possibility to use, in the near future,
the tunable filters opens a new path toward the performance of Cepheid
observations in the Walraven bands. 
Since new stellar
atmosphere model colours in the Walraven photometric system were available from
\citet{c99}, we have been able to calibrate the surface brightness 
function using $(V-B)$ and $(U-W)$ colours, in order to
apply the complete CORS method, including the $\Delta B$ term
(Eq.~\ref{eq-cors}).

The comparison of temperature curves for some Cepheids of our sample
with the temperature values obtained from spectroscopy has revealed
that the $(V-B)$ Walraven colour is a very good temperature indicator,
whereas a worse agreement is found for the gravity curves, probably due
to a lower accuracy in the measurement of the W band flux.

In this work we are able to accurately
calibrate the surface brightness and to derive the radius even for
Cepheids with $P\sim40$ days. 
The presence of long period stars in our sample allowed us to better
constrain the coefficients of the Period--Radius relation. 
In order to derive the Period--Radius relation, we have considered
only single stars, excluding binaries, since orbital motion and/or the
influence of the flux of a companion could affect the radius derivation. Furthermore we have
also excluded first overtone pulsators. Our final sample thus reduces 
to 27 Cepheids.

In order to take account of the uncertainty in the projection
factor value, we have considered two possible values, $p=1.27$ and
$p=1.36$, chosen in order to include the range of values found in
the literature. 

The Period--Radius relations obtained in the present work are the
following: 
\begin{eqnarray}
\log R=(0.75\pm 0.03)\log P+(1.10\pm0.03) \\
\log R=(0.75\pm0.03)\log P+(1.13\pm0.03)
\end{eqnarray}
for $p=1.27$ and $p=1.36$ respectively. We
excluded X Sct from the fit because of its large deviation
($>4\sigma$). We investigated the hypothesis that it can be an unknown
first overtone Cepheid  and our results suggest that it is a good
candidate, but further analysis is required.

The comparison with independent results in the literature has shown that the
agreement with the slope of our Period--Radius relations improves
including the $\Delta B$ term. The only exception is the
Period--Radius relation by 
\citet{rb97}, who find a flatter
slope using the surface brightness technique. Furthermore, the
uncertainties in our relations are of the same order of magnitude or
even smaller than those obtained using the infrared data.  
The zero--point of our relation decreases from $b=1.13$ to $b=1.10$ by
varying the projection factor from $p=1.36$ to $p=1.27$, providing a
better agreement with several authors. Therefore, we consider 
$\log R=(0.75\pm0.03)\log P+(1.10\pm0.03)$, obtained for $p=1.27$, as
our best fit Period--Radius relation. 

This result is also confirmed by the study of the Period--Luminosity
relation. The comparison with the results of other authors
\citep{gf98, kb04, fa07, bj03} shows that the agreement improves by decreasing the
projection factor values from 1.36 to 1.27. Thus we decided to keep
the value $p=1.27$ and considered the corresponding Period--Luminosity
relation as the best fit one: 
\begin{equation}
M_V=(-2.78\pm0.11)\log P -(1.42\pm0.11)
\end{equation} 
with a scatter of 0.13 mag which is smaller than other works.

In order to exclude possible uncertainty in the reddening, we also
derived the Period--Wesenheit relation. As expected from our findings
about the Period--Luminosity relation, the result obtained with
$p=1.36$ seems to make Cepheids too luminous with respect to other
results in the literature \citep{fa07, bm07}. Thus we decided to
keep the relation for $p=1.27$ again as the best fit one: 
\begin{equation}
W_{VI_C}=(-3.50\pm0.07)\log P -(2.72\pm0.07) .
\end{equation} 
which has a scatter of 0.09 mag.

Whereas the implications of the obtained relations for distance scale
applications and for the very important issue of the universality of
the Period--Luminosity relations deserve further investigation and
will be the subject of a forthcoming paper (Molinaro et al. in
preparation), it is worth noticing that the presented comparison with
independent results has provided a  test for the projection factor, 
suggesting a smaller value than the classical $p=1.36$. In
particular, $p=1.27$ appears to be the most accurate, in agreement with
the values obtained by \citet{g99,mk05,nm07}.  

\section*{Acknowledgments}
It is a pleasure to thank the referee Dr. P. Fouqu\'e, for his pertinent
concerns and constructive suggestions that helped us to improve both the
content and the readibility of the paper.

This work has benefitted from the use of the McMaster database
(http://crocus.physics.mcmaster.ca/Cepheid/) mantained by D. Welch.

\appendix

\section{The fit of theoretical grids}
\label{appA}
Here we show the result of the polynomial fit performed in order to
express the effective temperature, $\log T_{eff}$ and the effective
gravity, $\log g_{eff}$ as function of the $(V-B)$ and $(U-W)$ colours
and the Bolometric Corrections, $BC$, as function of temperature and
gravity:
\begin{eqnarray}
\log T_{eff}=a_0+a_1(V-B)+a_2(U-W)+\\
+a_3(V-B)(U-W) \nonumber 
\end{eqnarray} 
\begin{eqnarray}
\log g_{eff}=b_0+b_1(V-B)+b_2(U-W)^2+\\
b_3(V-B)(U-W)+b_4(V-B)^2(U-W)+ \nonumber \\
+b_5(V-B)(U-W)^2+ +b_6(V-B)^2(U-W)^2 \nonumber
\end{eqnarray} 
\begin{eqnarray}
BC=c_0+c_1(\log T_{eff})+c_2(\log T_{eff})^2+\\ 
c_3(\log g_{eff})+c_4(\log T_{eff})(\log g_{eff}) \nonumber   
\end{eqnarray}\label{eq-bc} 
The coefficients $a_i,\,b_i,\,c_i$ of the previous equations are
listed in Tab.~\ref{tab-coeff}. The rms of the previous relations are
0.0033 dex, 0.15 dex and 0.0032 respectively.

\begin{table*}
\begin{center}
\caption{Coefficients of the polynomial relations described in the Appendix ~\ref{appA}}
\begin{tabular}{c @{} c @{} c @{} c @{} c @{} c @{} c}
\hline
\hline
$a_0$ &\hspace{0.5 cm} $a_1$ &\hspace{0.5 cm} $a_2$ &\hspace{0.5 cm} $a_3$ &\hspace{0.5 cm} $a_4$ &\hspace{0.5 cm} $a_5$ &\hspace{0.5 cm} $a_6$\\
$3.959\pm0.004$ &\hspace{0.5 cm} $-0.666\pm0.012$ &\hspace{0.5 cm} $-0.251\pm0.013$ &\hspace{0.5 cm} $0.82\pm0.03$ & & &\\
\hline
$b_0$ &\hspace{0.5 cm} $b_1$ &\hspace{0.5 cm} $b_2$ &\hspace{0.5 cm} $b_3$ &\hspace{0.5 cm} $b_4$ &\hspace{0.5 cm} $b_5$ &\hspace{0.5 cm} $b_6$\\
$3.21\pm0.19$ &\hspace{0.5 cm} $23.0\pm3.0$ &\hspace{0.5 cm} $-24.0\pm2.0$ &\hspace{1
  cm} $-130.0\pm18.0$ &\hspace{0.5 cm} $59.0\pm12$ &\hspace{0.5 cm} $204.0\pm27$ &\hspace{0.5 cm} $-151.0\pm29$ \\
\hline 
$c_0$ &\hspace{0.5 cm} $c_1$ &\hspace{0.5 cm} $c_2$ &\hspace{0.5 cm} $c_3$ &\hspace{0.5 cm} $c_4$ &\hspace{0.5 cm} $c_5$ &\hspace{0.5 cm} $c_6$\\
$85.7\pm1.4$ &\hspace{0.5 cm} $-44.41\pm0.73$ &\hspace{0.5 cm} $5.75\pm0.10$ &\hspace{0.5 cm} $-0.12\pm0.02$ &\hspace{0.5 cm} $0.035\pm0.006$ & &\\
\hline
\end{tabular}
\label{tab-coeff}
\end{center}
\end{table*}

\section{The analysis with $(V-B)$-$(L-U)$ colours.}
\label{appB}
\begin{figure}
\includegraphics[width=84mm]{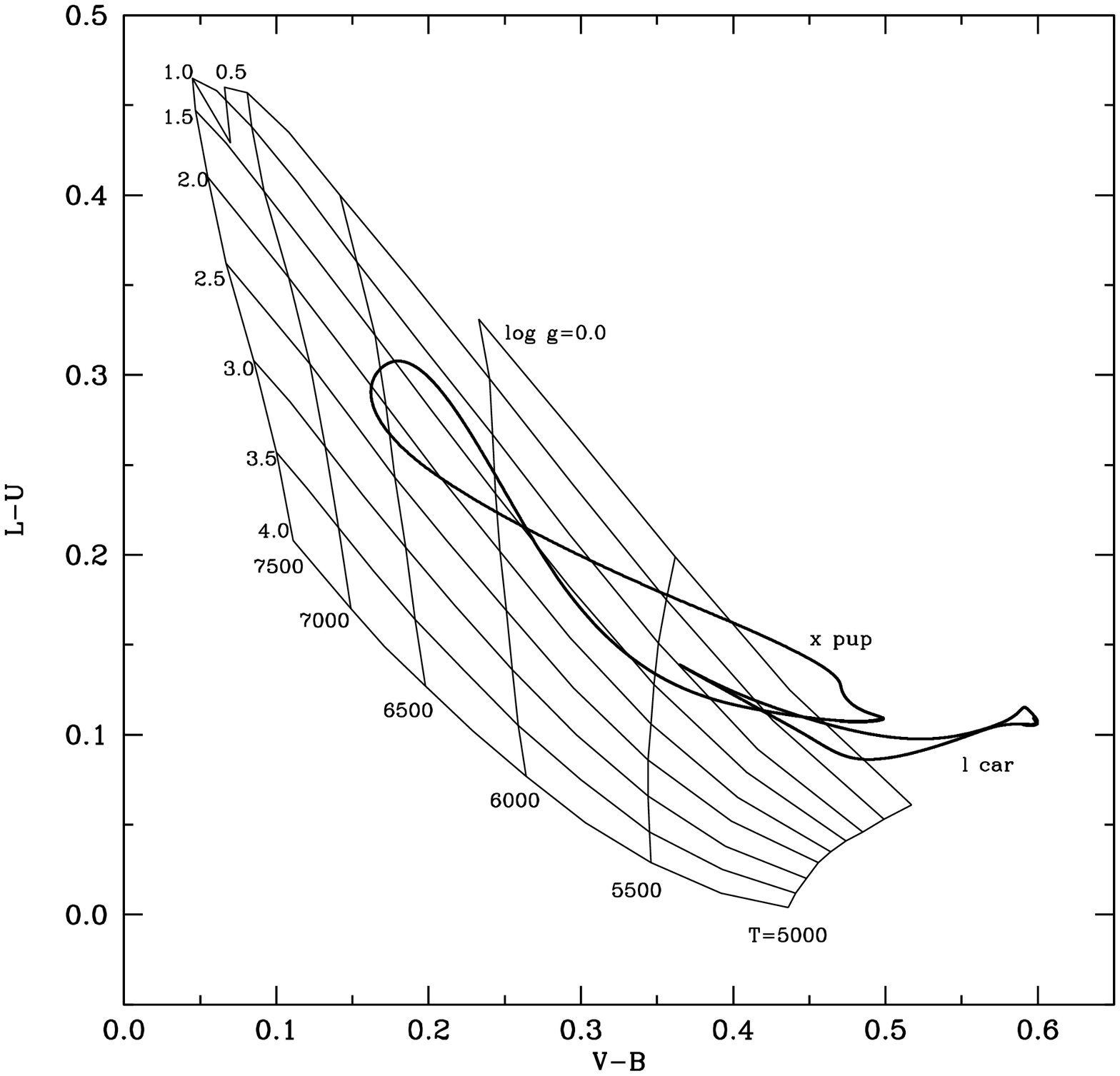}
\caption{Grids of models in the $(V-B)-(L-U)$ plane for the parameter
  range $5000\le T_{eff}<7500$ and $0.0\le \log g_{eff} \le 4.00$. The
coloured lines represent the loop of the two long period Cepheids l
Car ($P=35$ days) and x Pup ($P=26$ days).} 
\label{fig-griglia-lu-loop}
\end{figure}

Figure ~\ref{fig-griglia-lu-loop} shows the grid of models in the plane
$(V-B)-(L-U)$ for the parameter range $5000\le T_{eff}<7500$ and
$0.0\le \log g_{eff} \le 4.00$. According to the procedure described
in Appendix~\ref{appA}, we performed a fit of the grid using polynomial
equations in the variable $(V-B)$ and $(L-U)$. For brevity, we do not
give the explicit expressions of the fit, but we recall that the rms
of residuals around the fit are 0.0023 and 0.14 for $\log T_{eff}$ and
$\log g_{eff}$ respectively. 

In the same graphic the loops of the Cepheids l Car ($P=35$ days) and
X Pup ($P=26$ days) are also plotted. Since part of the  loops
falls outside of the grid, we are not able to accurately estimate the
structural parameters of the two stars for all phases of the
pulsational cycle. The same behaviour is observed for most Cepheids
with period longer than 11 days. It would be interesting to
investigate about this finding in a subsequent work. A possible reason
may be that the L band covers an extremely line-rich region of the
spectrum. Apart from a very large number of metallic lines, it
includes the higher Balmer lines and molecular lines of CN and CH. For
this reason, L may be more sensitive to dynamic effects (emission
components due to shockwaves, enhanced and time--variable
microturbulence) not included in the physics at the base of models,
hence more sensitive to breakdown of the quasi-hydrostatic equilibrium
assumption. The presence of molecular lines in L may also mean that
the Kurucz/Castelli model atmospheres have shortcomings at the cool
end of the grids that become particularly pronounced in L.

\begin{figure}
\includegraphics[width=84mm]{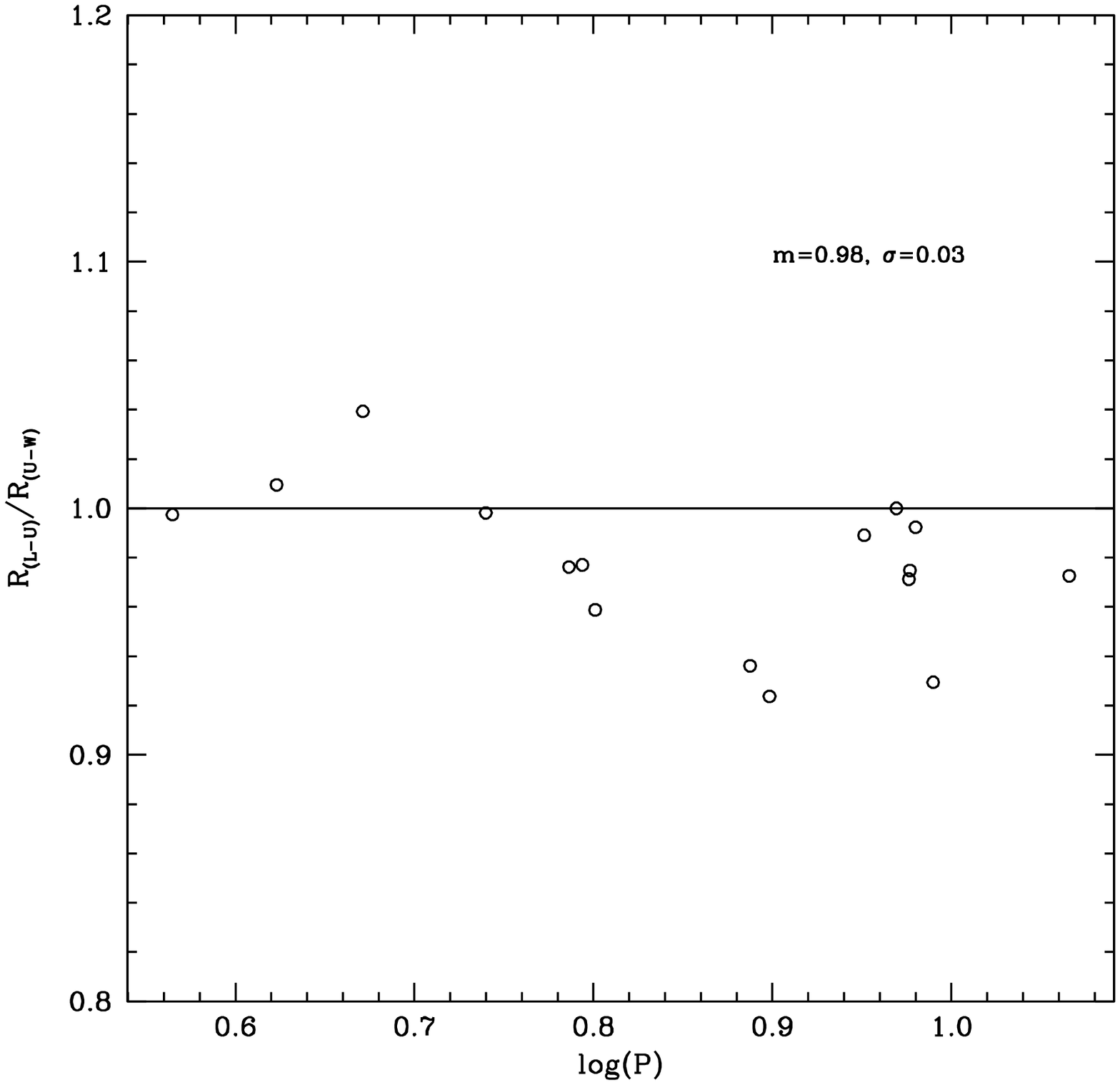}
\caption{The ratios between the new radii obtained with $(L-U)$ colour
  and the old values for a sample of singular Cepheids with
  colour-colour loop fully contained on the $(V-B)-(L-U)$ grid of
  models. The mean ratio and the standard deviation, indicated in red,
  are respectively  $m=0.98$ and $\sigma=0.03$.} 
\label{fig-ratios}
\end{figure}

\begin{figure}
\includegraphics[width=84mm]{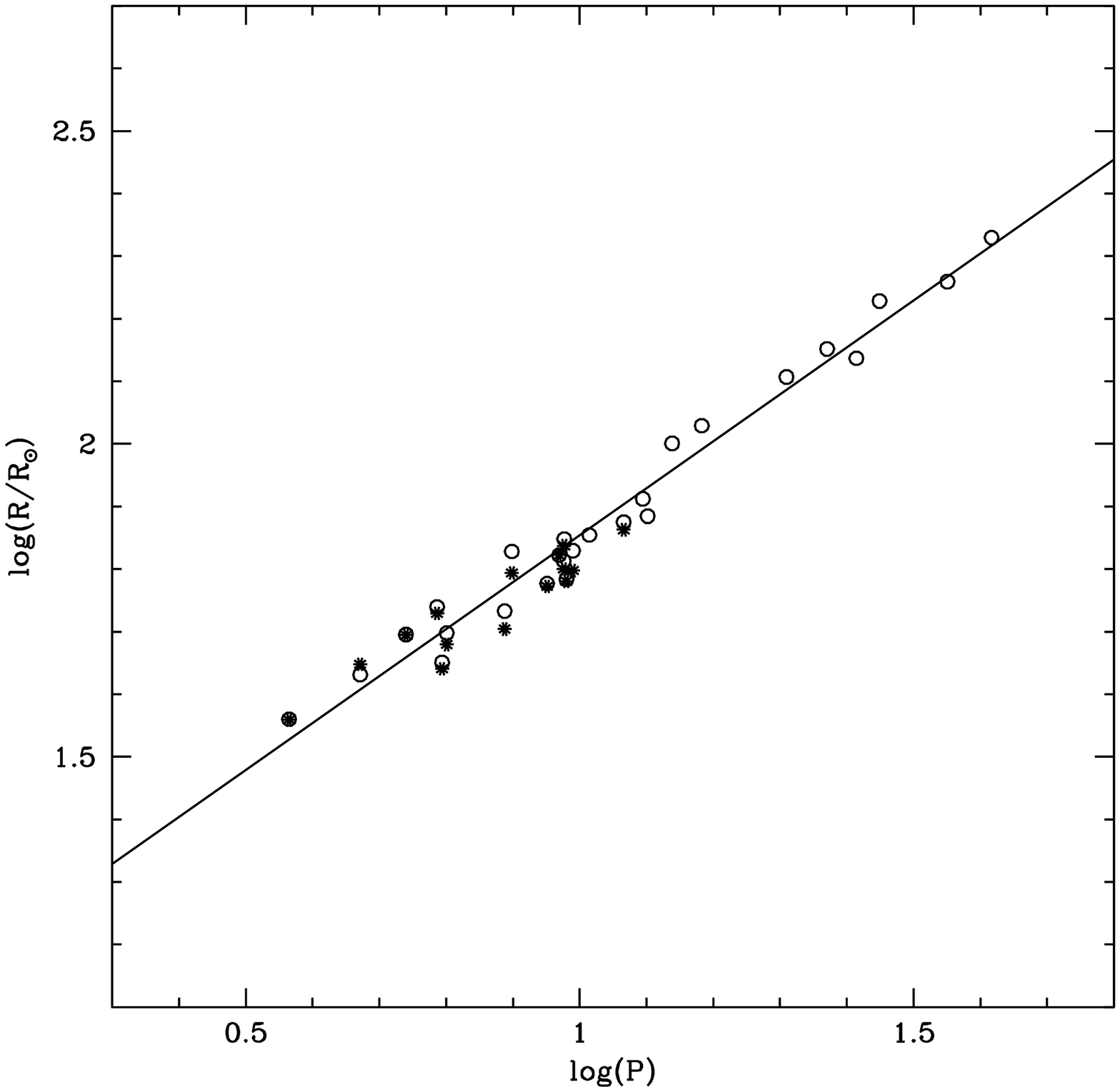}
\caption{The Period--Radius relation (Eq.~\ref{eq-pr-xsct-db}) is plotted
  together with the values of radius (blue stars) obtained with
  $(L-U)$ colour, for the sample of Cepheids having loop fully contained
  on the grid of $(V-B)-(L-U)$ models.} 
\label{fig-overplot}
\end{figure}

Using only the Cepheids with loops fully contained on the grids,
we have estimated their radii comparing them with old values
obtained using the $(U-W)$ colour. The ratios of their values, plotted
in Fig.~\ref{fig-ratios}, show that the differences are smaller than
$10\%$, with mean ratio equal to 0.98 and a standard deviation of 0.03. 
We have also overplotted the new radii on the Period--Radius relation
(Eq.~\ref{eq-pr-xsct-db}) and they seem to lie on the fitted relation as
shown in Fig.~\ref{fig-overplot}.  

On the basis of these analysis, we have decided to use the $(U-W)$
colour instead of $(L-U)$.

\end{document}